%% file: Combined_final_after_revision.tex
\documentclass[notitlepage,superscriptaddress,nofootinbib,amsmath,amssymb]{revtex4-1}

\usepackage[a4paper, margin=2.50cm]{geometry}

\usepackage{bbm}
\usepackage{graphicx}
\usepackage{fancyhdr, color}
\usepackage{hyperref}

\include{commands}

\begin{document}

\fancyhead[C]{\sc \color[rgb]{0.4,0.2,0.9}{Quantum Thermodynamics}}
\fancyhead[R]{}

\title{Single Particle Thermodynamics with Levitated Nanoparticles}

\author{James Millen}
\email{james.millen@kcl.ac.uk} 
\affiliation{Faculty of Physics, University of Vienna, Boltzmanngasse 5, Vienna 1090, Austria.}
\affiliation{Department of Physics, King's College London, Strand, London, WC2R 2LS, UK}
\author{Jan Gieseler}
\email{jgieseler@fas.harvard.edu} 
\affiliation{Department of Physics, Harvard University, 17 Oxford Street, Cambridge, MA 02138, USA.}

\date{\today}

\maketitle

\thispagestyle{fancy}

%%%%%%%%%%%%%%%%%%%%

%%%%%%%%%%%%%%%%%%%%%%%%%%%%%%%%%%%%%%%%%%%%%%%%%%%%%%%%%%%%%%%%%%%
This article appears as a chapter in the book `Thermodynamics in the Quantum Regime' (eds. Binder F., Correa L., Gogolin C., Anders J., Adesso G.), vol. 195 of the 'Fundamental Theories of Physics' series (Springer). 

\section*{Introduction}

The first demonstration of the stable 3D optical trapping of micron-scale particles was in the 1980s \cite{Ashkin1986}, and since then there has been an explosion of research using ``optical tweezers'', to the point that they are an off-the-shelf tool for physical and biological scientists. Using this system, it is possible to control and track the motion of mesoscopic objects with astounding precision. The first investigation of a microscopic thermodynamic process with an optically trapped particle was the realization of a Brownian ratchet \cite{Faucheux1995} and there was a strong increase of activity following the foundation of stochastic thermodynamics \cite{SiefertReview} and the discovery of fluctuation theorems such as the Jarzynski equality, with Seifert describing trapped colloidal particles as ``\emph{the} paradigm for the field (of stochastic thermodynamics)''.

So, what makes the trapped microparticle such a good platform for thermodynamic studies? First and foremost, its characteristic energy is comparable to that of the thermal fluctuations of the bath $\sim \kB \T{env}$. These small particles in harmonic optical potentials are simple, and considering only the centre-of-mass motion is for most cases sufficient to fully describe their behaviour\footnote{Recent work with levitated nanoparticles also considers rotational degrees of freedom \cite{Kuhn2017}.}. Having few degrees of freedom  enhances the relative role of thermal fluctuations via the central limit theorem: energy fluctuations of a system with $N$ degrees of freedom can be quantified by comparing the variance $\sigma^2 \propto N$ to the mean $\langle U \rangle \propto N$ of an extensive macroscopic quantity $U$, such as the total energy. For large $N$, $\langle U \rangle \gg \sigma$, whereas for small $N$, $\langle U \rangle \sim \sigma$ \cite{Gieseler2014}, illustrating the dominant role of fluctuations in systems with few degrees of freedom. 

Thus, with optical trap depths $>10^4\,$K and optical spring constants of $\sim \pN/\mum$, the motion of micron-sized particles is sensitive to thermal fluctuations, but not destructively so. 
The ability to dynamically alter the potential landscape in which the particle moves is also key to their application in studying thermodynamics. This can involve changing the depth of the optical potential, to realize compression stages in heat engines \cite{Schmiedl2008} or to speed-up equilibration \cite{Martinez2016}, or creating geometries with multiple stable trapping sites to test information thermodynamics \cite{Berut2012}. 

The majority of thermodynamic studies with optically trapped particles involve \emph{colloidal} particles: objects suspended in a liquid. In contrast, this chapter will consider \emph{levitated} nanoparticles, that is particles trapped in a gas or vacuum. It is somewhat experimentally more challenging than working in liquid, requiring deeper optical potentials due to reduced viscous damping, and loss of the particles from the trap at low pressures is a common problem.

Why work in this challenging regime at all, if the colloidal system has been so successful? Firstly, working in a gaseous environment gives us access to \emph{underdamped} dynamics, as opposed to the overdamped dynamics typically observed in a liquid.
The underdamped regime is of fundamental interest, since the inertia of a particle plays a role in the dynamics, whereas it can be mostly ignored in overdamped systems.
Secondly, and motivated by the subject matter of this book, the underdamped regime allows one to make the connection to the even more fundamental unitary evolution of quantum mechanical systems.
In addition, there is the potential to study quantum physics with these mesoscopic objects \cite{Chang2010}. The observation of quantum phenomena with levitated nanoparticles absolutely requires working in a good vacuum, since collisions with gas molecules cause rapid heating and decoherence. \\
%Therefore, future experiments with levitated nanoparticles in the quantum regime will help to characterize the sources of irreversibility in micro-engines and give new insight into the statistical properties of their efficiencies that could inspire new strategies in the design of efficient nano-motors.
%In addition, rapid progress in cooling the center of mass motion will enable the operation in the quantum regime, thereby realizing a textbook quantum Brownian particle.
%
%Secondly,

%One way to define the transition between these regions is to compare the harmonic frequency $\w$ of a trapped particle to the momentum damping rate $\g{CM}$, such that dynamics are underdamped when $\g{CM} \ll \w$. As an example, with typical $\w \sim 100\,$kHz, a 100\,nm radius silica nanosphere in a room temperature gas experiences $\g{CM} \sim\,$MHz at atmospheric pressures, and $\g{CM} \sim\,$mHz at $10^{-6}\,$mbar pressures. 

%\subsection{Scope of this chapter}

This chapter is intended as a pedagogical introduction to the dynamics of optically levitated nanoparticles with a focus on the study of single particle thermodynamics. Much of the work studying thermodynamics with nano- and micro-particles has taken place in liquid, and this chapter will avoid reviewing this impressive body of work, focussing instead on studies of thermodynamics with nanoparticles levitated in a gas. For a recent literature review we refer the reader to Ref. \cite{EntropyReview}. The authors will discuss extensions into the quantum regime where relevant throughout the chapter.

Section~\ref{sec:background} gives a detailed review of the stochastic and deterministic forces acting on an optically levitated nanoparticle, including a discussion of heating due to optical absorption. Section~\ref{sec:Brownian} describes the Brownian motion of a levitated particle, which will highlight the role of this system as a paradigm for studying stochastic thermodynamics. Section~\ref{sec:time_dep} will detail the utility of sculpting time-dependent potentials for trapped particles, in particular the ability to create effective baths and non-thermal states. Finally, section~\ref{sec:thermodynamics} will review and discuss recent experimental progress in realising important thermodynamics processes with levitated nanoparticles.

%%%%%%%%%%%%%%%%%%%%%%%%%%%%%%%%%%%%%%%%%%%%%%%%%%%%%%%%%%%

\section{The trapped nanoparticle system}
\label{sec:background}
A particle with radius $a \sim 100\,$nm has, generally speaking, of order $(a/a_o)^3\approx 10^{10}$ degrees of freedom, where $a_o$ is the size of the atoms making up the particle. However, for most practical purposes we characterize excitations within the particle by its internal temperature $\T{int}$ and the particle's external degrees of freedom, like its position $\mathbf{r}$, which describes its centre-of-mass motion, and its orientation.
In the context of single particle thermodynamics, the most relevant degree of freedom is the particle's position. Thus, we will focus our attention on this degree of freedom and only briefly mention the others in their relationship to the center-of-mass motion.  

%\subsection{Particle dynamics}
%For what follows, the centre-of-mass motion of a levitated nano-particle can be well described classically.
The equations of motion for the centre-of-mass can be well described classically and are given by Newton's second law

% =================================================
\begin{equation}
\ddot{\mathbf{r}}(t) \,+\, \boldsymbol{\g{CM}} \!\; \dot{\mathbf{r}} =\;  \frac{1}{m }\left[ \Ffluct(t) 
+ \mathbf{F}_{\rm det}(\mathbf{r}, t)
\right ],
\label{eq:NewtonLaw}
\end{equation}
% =================================================

where $m$ is the particle's and $\g{CM}$ the momentum damping rate, as discussed in detail below. We have isolated the contributions to the forces that act on the particle into stochastic forces $\Ffluct$ and deterministic forces $\mathbf{F}_{\rm det}$.

In the following section we discuss the origin of these forces and how they can be controlled in an experiment. This will lead to an effective description of the particle as a Brownian particle in a potential landscape, where the shape of the potential and the thermal bath can be controlled experimentally. This model is at the heart of many stochastic processes, which can therefore be simulated with this platform.

% =================================================
\begin{figure}[hbt]
  \includegraphics[width=0.9\textwidth]{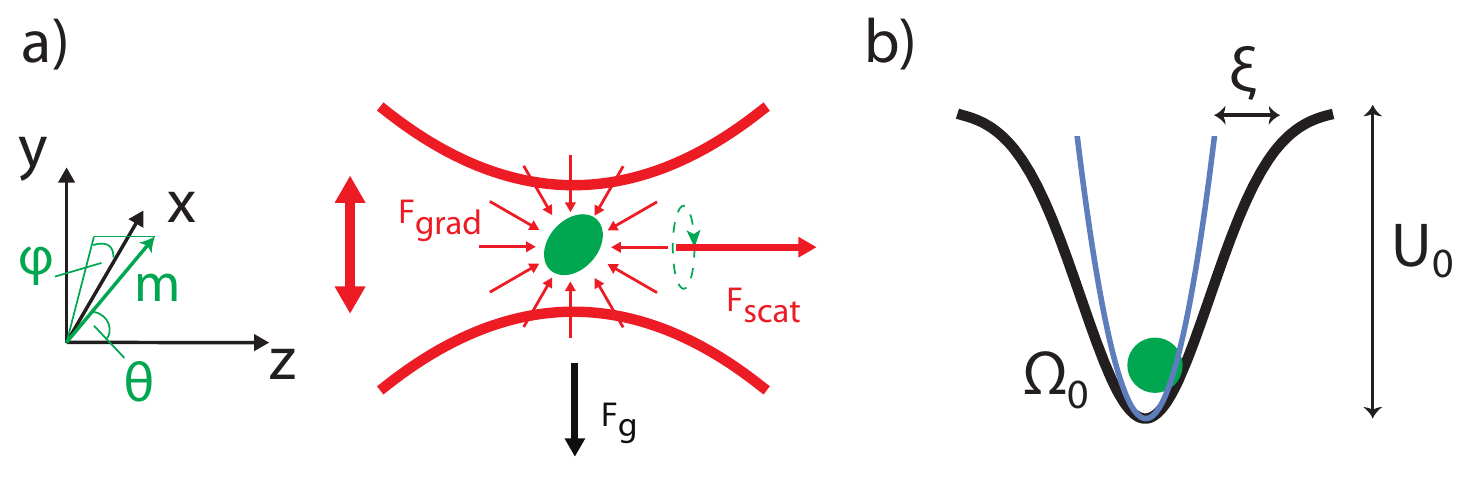}
  \caption{a.) An optical trap is formed by a tightly focused laser beam. In a single beam trap, the laser propagates along z. The non-conservative scattering force $\mathbf{F}_{\rm scat}$ acts along z while the gradient force points towards the maximum laser intensity, where the thick red lines on the figure represent the profile of a focussed light beam. In addition, the particle can experience torque and rotate in the trap. The orientation of a particle with a single symmetry axis is characterized by the two angles $\phi$ and $\theta$. The particle also feels a force $\mathbf{F}_{\rm g}$ due to gravity.
  b.) The gradient force forms an optical potential with depth $U_0$. The particle is trapped at the bottom of the potential and oscillates at frequency $\wo$. For large oscillation amplitudes, the nonlinearity of the potential is characterized by the Duffing parameter $\xi$.
   \label{fig:Schematic}}
\end{figure}
% =================================================

\subsection{Stochastic forces\label{sec:stochastic_force}}

The interaction of the particle with its environment has  mechanical (collisions with air molecules) and radiative (blackbody and scattering) contributions. These interactions lead to dissipation acting on the center-of-mass motion $\g{CM}^\noise$ and are the source of the random forces acting on the particle. The strength of the random forces is characterized by their power spectral densities $\Sff{\noise}$. For most practical purposes, they can be considered as frequency independent (white noise), that is the autocorrelation functions of the stochastic forces are $\langle \mathcal{F}_{\noise}(t) \mathcal{F}_{\noise}(t')\rangle =2\pi \Sff{\noise}\delta(t-t')$. After a time $\approx 1/\g{CM}$, where $\g{CM} = \sum_\noise\g{\noise}$ is the total damping rate, the center-of-mass motion of the particle reaches an effective thermal equilibrium, which is characterized by an effective temperature through the fluctuation-dissipation relation:

% =================================================
\begin{equation}
\label{eqn:temp_def}  
\T{CM} = \frac{\pi\Sff{} }{\kB m \g{CM} },
\end{equation}
% =================================================

where $\Sff{} = \sum_\noise\Sff{\noise}$ is the total force spectral density, $m$ the mass of the particle, and $\kB$ is Boltzmann's constant.
Below we describe the individual contributions. They are:
 collisions with air molecules ($\noise =$ gas), radiation damping ($\noise =$ rad), feedback or cavity damping ($\noise =$ fb), stochastic driving ($\noise =$ drive), and in the quantum regime noise driving wavefunction collapse ($\noise =$ CSL).

\subsubsection{Gas damping}

% #### GAS DAMPING ###
% ######################
For pressures higher than $\sim 10^{-6}\,\rm mbar$, the dominant contribution to the stochastic forces is due to collisions with surrounding air molecules, and the damping rate is given by \cite{Beresnev:1990tv} 

% =================================================
\begin{equation}\label{eq:gas_damping}
  \frac{\g{gas}}{2\pi} = 3\viscosity \frac{a}{m}\frac{0.619}{0.619+\Kn}\left(1+c_K\right),
  \quad
  \Sff{gas} = \frac{m\kB \T{gas}}{\pi}\g{gas},
\end{equation}
% =================================================

where $c_K = 0.31 \Kn\left/\left(0.785+1.152\Kn+\Kn^2\right)\right.$, $\viscosity$ is the viscosity coefficient, which for a dilute gas is $\viscosity = 2\sqrt{\mgas \kB\T{gas}}\left/ 3\sqrt{\pi} \crosssectiongas \right.$ and $\Kn = \bar{l}/a$ is the Knudsen number for the free mean path $\bar{l} = \kB \T{gas}\left/(\sqrt{2}\crosssectiongas\Pg)\right.$, $\crosssectiongas = \pi \dm^2$, $\dm = 0.372\,\rm nm$ is the diameter of the air molecules and $\mgas$ their mass.

% #### HIGH DAMPING ####
% ######################
For high pressures (where $\Kn\ll 1$), the interaction with the gas is so strong that the particle motion is heavily damped and its internal temperature $\T{int}$ and centre-of-mass temperature $\T{CM}$ quickly thermalize with the gas temperature $\T{gas}$. In this regime, the damping becomes independent of pressure $\g{CM}/2\pi\approx 3 a \viscosity/m$, as predicted by Stokes' law.

% #### TWO GAS MODEL ###
% ######################
For decreasing pressure, the mean free path of the gas molecules increases (e.g. $\bar{l}\sim 60\,\mu$m at 1\,mbar). As a consequence, the particle no longer thermalizes with the gas since the impinging gas molecules no longer carry away enough thermal power to balance the optical absorption from the trapping laser. Due to the increased internal temperature $\T{int}$ of the particle, the average energy of the gas molecule after a collision with the particle increases. The process by which a surface exchanges thermal energy with a gas is called accommodation, which is characterized by the accommodation coefficient,

% =================================================
\begin{equation}
\label{eqn:accomm}
 \acccoeff = \frac{\T{em} - \T{gas}}{\T{int} - \T{gas}},
\end{equation} 
% =================================================

where $\T{int}$ is the temperature of the surface, $\T{gas}$ the temperature of the impinging gas molecules and $\T{em}$ the temperature of the gas molecules emitted from the surface. Accommodation quantifies the fraction of the thermal energy that the colliding gas molecule removes from the surface, such that $\acccoeff =1$ means that the molecule fully thermalizes with the surface. Since the mean free path in a dilute gas is long, one can safely assume that a molecule that comes from the particle surface will not interact again with the particle before thermalizing with the environment. Consequently, we can consider the particles that impinge on the particle surface and those that leave the surface as being in equilibrium with two different baths with temperatures corresponding to the temperature of the environment and the particle surface, respectively. Therefore, we get an additional contribution to the damping from the emerging hot molecules

% =================================================
\begin{equation}\label{eq:gas_damping_2bath}
  \frac{\g{em}}{2\pi} = \frac{1}{16}\sqrt{\frac{\T{em}}{\T{gas}}}\g{gas} ,\quad
  \Sff{em} =
 \frac{m\kB}{\pi}\left[\acccoeff\T{int} +(1-\acccoeff\right)\T{gas}]\g{em},
\end{equation}
% =================================================

as experimentally observed by Millen \emph{et al.} \cite{Millen2014}. Note that $\T{em}$ can be calculated using eqn.~\eqref{eqn:accomm}. In addition to this noise contribution, the internal temperature of the particle can also cause deterministic forces to act on the particle's centre-of-mass motion through the photophoretic effect, where absorbing particles are repelled from the optical intensity maxima \cite{Lewittes1982}. However, since photophoretic forces require a temperature gradient across the particle, they vanish for sub-wavelength particles, which are mostly used in vacuum trapping experiments (typical trapping laser wavelengths range from 532\,nm to 1550\,nm). 

% #### LOW DAMPING REGIME ###
% ###########################
For pressures below $\Pg =0.57 \kB \T{gas} \left/\crosssectiongas a\right.\approx 54.4\,{\rm mbar} \times (a/ \viscosity {\rm m})^{-1}$, where the mean free path is much larger than the radius of the particle ($\Kn\gg 1$), the damping becomes linear in pressure

% =================================================
\begin{equation}\label{eq:gas_damping_lin}
% \frac{\g{gas}}{2\pi} =  \frac{3(8+\pi\sqrt{\T{em}/\T{gas}} )}{8\pi\sqrt{2}}\frac{\viscosity\crosssectiongas}{\kB \T{gas} \rho}\frac{\Pg}{a} % Epstein
% = \cg \frac{\Pg}{a},
\frac{\g{gas}}{2\pi} =  \frac{3}{\pi\sqrt{2}}\frac{\viscosity\crosssectiongas}{\kB \T{gas} \rho}\frac{\Pg}{a}, % Epstein
\end{equation}
% =================================================

where $\rho$ is the density of the particle. The two expressions eqns.~\eqref{eq:gas_damping} \& \eqref{eq:gas_damping_lin} differ by less than 10\% for $\Kn\gg 1$, with the discrepancy due to numerical accuracy when calculating the constant factors in eqn.~\eqref{eq:gas_damping} \cite{Beresnev:1990tv}. The total damping due to the hot particle with the gas environment is $\g{em}+\g{gas} =2\pi \cg \Pg / a$, where typically $\cg\approx   50\, \Hz (\mum/ \mbar)$. 
When considering operation in the quantum regime, it is absolutely necessary to work under extremely good vacuum conditions, as collisions with gas molecules cause rapid decoherence and heating out of the ground state.
As an example, a 100\,nm radius silica nanosphere in a room temperature gas experiences $\g{gas} \sim\,$MHz at atmospheric pressures, and $\g{gas} \sim\,$mHz at $10^{-6}\,$mbar pressures.

So far, we have considered spherical particles with translational degrees of freedom. However, in general the particle has some anisotropy and is free to rotate within the trap. The orientation of the particle with respect to the trap (see fig.~\ref{fig:Schematic}) is described by the angles $(\phi, \theta)$, where $\phi$ is the angle between the $x$-axis and the projection onto the $x-y$ plane, and $\theta$ is the angle between the particle axis and the $z$-axis. The particle axis is usually defined along its symmetry axis and represented here by the vector $\mathbf{m}$. For anisotropic particles, e.g. a cylinder, the friction term is different along each of the axes, and depends upon the alignment $\mathbf{m}$ of the particle. As a consequence, the friction coefficient has to be replaced by a tensor $\boldsymbol{\Gamma}$ and the damping in a direction $\mathbf{s}$ is given by $\boldsymbol{\Gamma}\cdot \mathbf{s}$. In the low pressure regime, the friction tensor of the translational degrees of freedom for a particle with a single symmetry axis can be derived analytically \cite{Martinetz2018}. As an example, for a cylinder of diameter $d$

% =================================================
\begin{equation}
\label{eqn:cylinder_damp}
 \frac{\boldsymbol{\Gamma}_\text{trans}}{2\pi} = 
 6\sqrt{2}\frac{\viscosity\crosssectiongas}{\kB \T{gas} \rho}\frac{\Pg}{\rodd} 
 \left (2 - \frac{1}{2}\acccoeff + \frac{\pi}{4}\acccoeff \right )
 \left ( \1 - \frac{8 -6\acccoeff +\pi\acccoeff}{8 -2\acccoeff +\pi\acccoeff}\mathbf{m}\otimes\mathbf{m}\right ).
\end{equation} 
% =================================================

For the rotational degrees of freedom we find that the damping is isotropic and given by

\begin{subequations}
% =================================================
\begin{equation}
\label{eqn:sphere_rot_damp}
 \frac{\g{rot}^\text{sphere}}{2\pi} =  \frac{30 \acccoeff}{8\pi\sqrt{2}}\frac{\viscosity\crosssectiongas}{\kB \T{gas} \rho}\frac{\Pg}{a}, % Epstein
\end{equation}
% =================================================

for a sphere and 

% =================================================
\begin{equation}
\label{eqn:cylinder_rot_damp}
\frac{\g{rot}^\text{cyl}}{2\pi} = 
 6\sqrt{2}\frac{\viscosity\crosssectiongas}{\kB \T{gas} \rho}\frac{\Pg}{\rodd} 
 \left (2 - \frac{1}{2}\acccoeff + \frac{\pi}{4}\acccoeff \right ),
\end{equation}
% =================================================
for a cylinder. 
\end{subequations}

\subsubsection{Noise from optical fields}
\label{sec:photon}

% #### SHOT NOISE ###########
% ###########################
At very low pressure ($\leq 10^{-6}\,\rm mbar$), gas damping rates become extremely small and photon shot noise starts to dominate \cite{Jain2016a}. Photon shot noise is a consequence of the particulate nature of light. Photons arrive at discrete times, where the number of photons arriving per time interval $\Delta t$ is given by $\sqrt{\Delta t \Popt\left/\hbar \wopt \right.} $, where $\Popt$ and $\wopt$ are the optical power and frequency, respectively. The recoil from the fluctuating number of phonons impinging on the nanoparticle can be modelled as an effective bath with the characteristics

% =================================================
\begin{equation}\label{eq:radiation_damping}
  \frac{\g{rad}}{2\pi} = \dpcoeff\frac{\Pscat}{2\pi m c^2}
\quad{\rm and}\quad
  \Sff{rad} = \dpcoeff \frac{\hbar \omega \Pscat}{2\pi c^2},   
\end{equation}
% =================================================

where $\dpcoeff$ depends on the direction of motion of the particle with respect to the polarization of the laser and is $\dpcoeff=2/5$ for motion along the direction of polarization and $\dpcoeff=4/5$ for motion perpendicular to the polarization. The scattered power is $\Pscat = \scatcross \Iopt$, where $\scatcross = |\alpha|^2\kopt^4/6\pi\epsilon_0^2$ with $\alpha$ the particle polarizability, and $\Iopt$ is the laser intensity. The effective temperature of this bath can be calculated via eqn.~\eqref{eqn:temp_def}.

The noise processes described so far are present in any experiment with optically levitated nanoparticles in high vacuum. In addition, random forces and damping can be introduced through external fields that are under experimental control. Importantly, since energy can be injected or extracted from the particle, i.e. it is not in a thermal equilibrium, the fluctuation-dissipation relation does not have to hold and the effective damping and temperatures can be controlled independently.

For instance, by parametric feedback damping (see also section~\ref{sec:non-thermal}), the temperature alone is not sufficient to give a full description of the bath. Ideal feedback cooling damps the particle motion at a rate $\g{fb}$ without adding any fluctuating forces, thus $\Sff{fb}=0$ and it is therefore referred to as cold damping. 
Similarly, cavity cooling up-converts the particle energy to optical frequencies, which are effectively at zero temperature because $\hbar \omega \gg \kB \T{env}$ in a room temperature environment. Conversely, fluctuations of the trapping or additional control fields only add fluctuating forces without providing damping. Hence, $\g{drive} = 0$ and $\Sff{drive} = \q^2 S_{\rm qq}$, where $\q$ is the coupling parameter to the control field and $S_{\rm qq}$ its spectral density.
This can be realized for example with fluctuating electric fields, where $\q$ corresponds to the charge on the particle. 
A real feedback signal is noisy, since any measurement is accompanied by noise, and this will heat the particle motion without providing additional damping. In addition, one has to consider correlations between the measurement signal and the particle motion when interpreting the measurement result, in particular when using a linear feedback signal.

\subsubsection{Noise due to wavefunction collapse}

There are a class of theories, called \emph{collapse models}, which aim to phenomenologically explain why we do not observe superposition states of macroscopic objects \cite{BassiReview}. These models invoke a (classical) noise field, which acts upon particles in a mass-dependent way, to ensure localization of the wavefunction. There are various proposed forms of the noise, including white noise fields which violate conservation of energy, to coloured and dissipative noise with a finite temperature (suggested to be between 0.1-10\,K). The noise induces a type of Brownian motion on the centre-of-mass or alignment of the particle, which in principle can be observed. 
%To illustrate the nature of this process, first consider the random force $\mathcal{F}_{\rm gas}(t)$ exerted on a particle due to collisions with gas molecules (Brownian motion), which is quantified as $\langle \mathcal{F}_{\rm gas}(t) \mathcal{F}_{\rm gas}(t')\rangle = 2\pi\Sff{gas}\delta(t-t')$, where $\Sff{gas}$ is defined in eqn.~\eqref{eq:gas_damping_2bath}.
%Equivalently, for a model of wavefunction collapse known as Continuous Spontaneous Localization (CSL), the random force $\mathcal{F}_{\rm CSL}(t)$ is defined through $\langle \mathcal{F}_{\rm CSL}(t) \mathcal{F}_{\rm CSL}(t')\rangle = \Sff{CSL}\delta(t-t')$, where $\Sff{CSL} = \lambda_{\rm CSL}(\hbar/r_{\rm CSL})^2\alpha_{\rm CSL}$. The parameters $\lambda_{\rm CSL}$ and $r_{\rm CSL}$ are a phenomenological rate and length scale, respectively.
%The factor $\alpha_{\rm CSL}$ is mass and geometry dependent, and as an example is proportional to the mass $m^{2/3}$ for a sphere. For a thorough discussion of this process, see the review by Bassi \emph{et al.} \cite{BassiReview}. 
For example, a model of wavefunction collapse known as Continuous Spontaneous Localization (CSL), predicts a random force with $\Sff{CSL} = \lambda_{\rm CSL}(\hbar/r_{\rm CSL})^2\alpha_{\rm CSL}$, where the parameters $\lambda_{\rm CSL}$ and $r_{\rm CSL}$ are a phenomenological rate and length scale, respectively.
The factor $\alpha_{\rm CSL}$ is mass and geometry dependent, and as an example is proportional to the mass $m^{2/3}$ for a sphere. For a thorough discussion of this process, see the review by Bassi \emph{et al.} \cite{BassiReview}.

\subsection{Deterministic forces}
In addition to stochastic forces, which we described in section~\ref{sec:stochastic_force}, the particle is also subject to deterministic forces. They are gravity $\F{g} = m \mathbf{g}$, electric forces $\F{e} = q\mathbf{E}$ if the particle carries a charge $q$, magnetic forces $\F{mag} = \nabla(\boldsymbol{\mu}\cdot\mathbf{B})$ if the particle has  a magnetic dipole moment $\boldsymbol{\mu}$ and optical forces $\F{opt}$. Most experiments with levitated particles in vacuum use optical forces to create a stable trap. This gives a great deal of flexibility since optical fields can be controlled very well in both intensity and position, allowing the creation of almost arbitrary fluctuating force fields. Particles that are much smaller than the wavelength $\lambdaopt$ of the trapping laser $a \kopt \ll 1$, where $\kopt = 2\pi/\lambdaopt$, can be treated as dipoles in the Rayleigh approximation. The polarizability for a particle with volume $V$ is thereby given by

% =================================================
\begin{equation}
\label{eqn:polarisability}
\polar{0} = \eo V \chitensor,
\end{equation}
% =================================================

where the total susceptibility of the particle  $\chitensor = \chitensor_e\left(1+\depol \chitensor_e\right)^{-1}$ depends on the material via the material susceptibility $\chitensor_e$ and on its geometry through the depolarization tensor $\depol$, which in general are both rank-2 tensors. However, for isotropic materials, the material susceptibility simplifies to a scalar $\chi_e$ and similarly for a sphere the depolarization tensor is isotropic and simplifies to a scalar $N = 1/3$. Thus, for a sphere we recover the Clausius-Mossotti relation $\chi = 3(\epsilon_p-1)/(\epsilon_p+2)$, where we use $\epsilon_p= 1+\chi_e$.

For a particle with a uniaxial anisotropy, e.g. a cylinder, the susceptibility $\chitensor = \diag(\chi_\|, \chi_\perp, \chi_\perp)$, has a component $\chi_\parallel$ parallel and a component $\chi_\perp$ perpendicular to the symmetry axis. For example, the depolarization tensor of a cylinder is $\depol = \diag(0, 1/2,1/2)$ in the frame of the cylinder, where the cylinder axis is along the $x$-axis. Consequently, $\chi_{\|} = \er - 1$, $\chi_{\perp} = 2(\er - 1)/(\er +1)$ for a cylinder with isotropic $\er$.
This means the maximal polarizability of a cylinder is $(\er+2)/3$ times higher than for a sphere of the equivalent volume. For silica, this is a factor of 2, whereas it is a factor of 4.6 for silicon.

In general, the particle reacts to the total field, that is the sum of the incident and the scattered field. The total field is the self consistent solution to Maxwell's equations and has to be calculated generally with numerical methods. 
However, for a spherical particle, the modified polarizability

% =================================================
\begin{equation}\label{eq:polarisability}
  \polar{} = \polar{0}\left(1-i\frac{\kopt^3\alpha_0}{6\pi\epsilon_0}\right)^{-1},
\end{equation}
% =================================================
accounts for the radiation reaction of the particle to its own scattered field, such that the induced polarization due to a field $\mathbf{E}_0$ is $\mathbf{P} = \alpha \mathbf{E}_0$. We introduce $\alpha'$ and $\alpha''$ to refer to the real and imaginary part of the polarisability, respectively.

Knowing the polarizability, we can calculate the optical force for sub-wavelength particles in the Rayleigh approximation. The optical force has conservative and non-conservative contributions

% =================================================
\begin{equation}\label{eq:Fopt}
  \F{opt} = 
\alpha'\nabla I_0/4 +\totcross\left[
\mathbf{S}/c+
 c \nabla\times \mathbf{L}\right],
\end{equation}
% =================================================

where the total cross-section $\totcross =\alpha''\kopt/\epsilon_0$ is the sum of the absorption and scattering cross-sections. The optical intensity at the field maximum $I_0$ is related to the field $E_0$ through $I_0 = c\epsilon_0E_0^2/4$.  The first term in eqn.~\eqref{eq:Fopt} is a conservative force. It pulls particles with a high refractive index relative to their surroundings toward the region of maximum light intensity.
In an optical tweezers, this is the focal volume of the light beam.
 
The second term is the non-conservative scattering force, which has two contributions: the radiation pressure term, which is proportional to the time averaged Poynting vector $\mathbf{S}=\left<\mathbf{E}\times\mathbf{H}^*\right>$, $\mathbf{H}$ being the magnetic field, and a curl force associated to the non-uniform distribution of the time averaged spin density of the light field $\mathbf{L} = -i\epsilon_0\left<\mathbf{E}\times\mathbf{E}^*\right>\left/4\wopt\right.$, $\langle\dots\rangle$ being a time average. The curl force is zero for a plane wave but can be significant for a tightly focused beam in an optical tweezers\footnote{For a Gaussian beam with waist $\waist_0$, we estimate the curl force as $F_{\rm curl} \approx -2 F_{\rm scat}\left/\waist_0^2 \kopt^2\right.$, where $F_{\rm scat} \sim \sigma_{\rm tot}\Popt\left/\waist_0^2 c\right.$. 
}. However, since $\alpha''/\alpha'\propto a^3$, the non-conservative forces vanish for small particles and we will neglect them in the following discussion.

\subsubsection{Optical potential}
The conservative force in eqn.~\eqref{eq:Fopt} can be expressed as the gradient of a potential $\F{opt} = -\nabla U_{\rm opt}$. Even for a tightly focused laser beam, the optical intensity distribution is to a good approximation described by a transverse Gaussian profile and a Lorentzian profile along the direction of beam propagation.
Thus, for a single focused laser beam the optical potential reads

% =================================================
\begin{equation}\label{eq:potential}
U_{\rm opt}(\mathbf{r}) =  
\frac{-\Uopt{}}{1+(z/z_0)^2}\exp\left[
   -\frac{2}{1+(z/z_0)^2}\left(
   	\frac{x^2}{\waist_x^2}+
   	\frac{y^2}{\waist_y^2}
   	\right)\right],
\end{equation}
% =================================================

where $\Uopt = \alpha' E_0^2\left/4\right.$ is the potential depth, $\mathbf{r}$ is the position of the particle, $\waist_x$, $\waist_y$ denote the transverse extent of the focus, and we define a longitudinal waist $\waist_z$ via the Rayleigh range $z_0 = \waist_z/\sqrt{2} \approx \pi\waist_0^2/\lambdaopt$, which gives the depth of focus. Note that for tightly focused laser beams, as are commonly used in optical trapping, the field distribution is slightly elongated along the direction of polarization of the incident field.
Integrating the Poynting vector over a cross-section that is transverse to the direction of propagation allows us to relate the field intensity at the centre of the focus $E_0^2$ to the optical power of the trapping laser
$
  \Popt = \int_s \left<\mathbf{S}\right>\mathbf{\d s} = \pi c\epsilon_0 \waist_0^2 E_0^2/4
$, where $ \waist_0^2 =  \waist_y  \waist_x$. At the bottom of the potential, the centre-of-mass motion is harmonic, with frequencies

% =================================================
\begin{eqnarray}\label{eqn:frequencies_translational}
\w_{i} &= &2 \sqrt{\frac{\chi}{c\pi\rho }}\frac{\sqrt{\Popt}}{\waist_0 \waist_i}, 
\end{eqnarray} 
% =================================================

along the three directions ($q = x, y, z$). For larger oscillation amplitudes, the motion becomes anharmonic, and the nonlinear coefficients can be obtained from higher derivatives of the optical potential (see section~\ref{sec:non-lin-BM}).

\subsubsection{Rotation}

As we have already seen in section~\ref{sec:stochastic_force}, the light matter interaction is more complicated for anisotropic particles, since it depends upon the alignment of the object relative to the polarization axis of the field\footnote{For a thorough treatment, including optical scattering, see \cite{Kuhn2017} and references therein.}. 
The induced polarization is $\mathbf{P} = \boldsymbol{\alpha}\mathbf{E}$.
Consequently, the particle experiences an optical torque
% =================================================
\begin{equation}
  \mathbf{N}_\text{opt} = \langle \mathbf{P} \times \mathbf{E}^*\rangle,
\end{equation}
% =================================================
which aligns the particle along the polarization axis. For small deflections from the polarization axis the angular motion is harmonic. For a cylinder of length $l$ the frequencies are

% =================================================
\begin{equation}\label{eqn:frequencies_rotational}
\Omega_{\azimuthalangle} = \sqrt{\frac{24 \Popt\chi_{\|}}{ \pi \rho c \waist_0^2 \rodl^2} \left( \frac{\Delta\chi}{\chi_{\|}} + \frac{(\kopt \rodl)^2}{12}\right)},  \quad
\Omega_{\polarangle}=\sqrt{\frac{24 \Popt \Delta\chi}{ \pi \rho c \waist_0^2 \rodl^2}},
\end{equation} 
% =================================================

where the term $(\kopt l)^2/12$ is a correction term that accounts for the particle's finite
extension.

In contrast to linearly polarized light, the polarization axis of circularly polarized light rotates at the optical frequency. This is too fast for the particle to follow. Nonetheless, light scattering transfers the angular momentum of the light to a particle with polarization anisotropy, which can originate from the intrinsic birefringence of the particle or from the anisotropic shape of the particle (c.f. eqn.~\eqref{eqn:polarisability}). The torque that results from angular momentum transfer is for a cylinder \cite{Kuhn2017}:

% =================================================
\begin{equation}
N_\polarangle= \frac{\Delta\chi \rodl^2 \rodd^4 \kopt^3}{96 c  \waist_{0}^2}  \left [\Delta \chi \eta_1(\kopt \rodl) +\chi_\perp \eta_2(\kopt \rodl) \right ] \Popt,
\end{equation}  
% =================================================

where the functions $\eta_{1,2}(\kopt \rodl)$ are given by

% =================================================
\begin{eqnarray}
\eta_1(\kopt \rodl) & = & \frac{3}{4} \int_{-1}^1 \mathrm{d}s~(1 - s^2) \mathrm{sinc}^2 \left ( \frac{\kopt \rodl s}{2} \right ), \notag\\
\eta_2(\kopt \rodl) & = & \frac{3}{8} \int_{-1}^1 \mathrm{d}s~(1 - 3s^2) \mathrm{sinc}^2 \left ( \frac{\kopt \rodl s}{2} \right ).
\end{eqnarray}
% =================================================

For short rods, $\kopt \rodl \ll 1$, one has $\eta_1 \simeq 1$ while $\eta_2 \simeq 0$. The rotational frequency $\Omega_{\rm rot}$ is given by the balance between the torque $N_\polarangle$ and the damping $\g{rot}^{\rm cyl}$, such that $\Omega_{\rm rot} = N_\polarangle/(I\g{rot}^{\rm cyl})$, where $I$ is the moment of inertia $I = m\rodl^2/12$. We note that this analysis is true in the Rayleigh-Gans approximation, where
$\kopt\rodl(\er -1)\ll 1$ and $\pi \kopt^2 d^2(\er -1)\ll 1$.

\subsection{Internal temperature}
\label{sec:internalT}
In the previous section we discussed the behaviour of a hot sphere levitated in gas and how its internal temperature $\T{int}$ couples to the centre-of-mass motion, which we characterized by its centre-of-mass temperature $\T{CM}$. In this section we will consider the process by which an optically levitated nanoparticle heats up.

Following Bateman \emph{et al.} \cite{Bateman2014} and Chang \emph{et al.} \cite{Chang2010}, the interaction between a sub-wavelength ($a<\lambdaopt$) sphere of radius $a$ and a light field of frequency $\omL$, is governed by the complex polarisability $\polar{}$.
The frequency dependent permittivity is related to the complex refractive index through $\eom{} = n(\omega)^2$. While $\polar{0}'$ determines the optical potential, $\polar{0}''$ determines optical absorption, with absorption cross-section $\abscross = \polar{0}'' \kopt/\eo$.
The bulk temperature depends on several competing processes: heating through absorption of the trapping light $\omL$, optical absorption of blackbody radiation with a spectral absorption rate $\rho_{\rm abs}$, and cooling through blackbody emission at a spectral emission rate $\rho_{\rm emis}$ and through energy exchange with the background gas. The blackbody spectral rates are given by:

% =================================================
\begin{eqnarray}
\label{eqn:absemis}
\rho_{\rm abs}(\ombb) &=& \frac{(\ombb/(\pi c))^2\abscross(\ombb)}{\exp(\hbar\ombb/(\kB \T{env}))-1}, \notag \\
\rho_{\rm emis}(\ombb, \T{int}) &=& \left ( \frac{\ombb}{\pi c}\right )^2 \abscross(\ombb) \exp \left (-\frac{\hbar\ombb}{\kB \T{int}}\right ),
\end{eqnarray}
% =================================================

where $\T{env}$ is the ambient temperature of the environment, and $\T{int}$ the surface temperature of the sphere (which we assume is equal to the bulk temperature). Following Chang \emph{et al.} \cite{Chang2010}, we can integrate across the blackbody spectrum to find the rate at which the sphere absorbs or emits blackbody energy:

% =================================================
\begin{eqnarray}
\label{eqn:bbenergy}
 \dot{E}_{\mathrm{abs}}^{\mathrm{bb}} &= \frac{24 \xi_R(5)}{\pi^2\eo c^3 \hbar^4} \alpha_{\rm bb}'' (\kB \T{env})^5 \notag \\
 \dot{E}_{\mathrm{emis}}^{\mathrm{bb}} &= -\frac{24 \xi_R(5)}{\pi^2\eo c^3 \hbar^4} \alpha_{\rm bb}'' (\kB \T{int})^5,
\end{eqnarray}
% =================================================

where $\xi_R(5) \approx 1.04$ is the Riemann zeta function, and $\polar{bb}$ is averaged over the blackbody spectrum, such that for silica, which is the material most commonly used in optical levitation experiments, $\polar{bb}'' \approx 4\pi\eo a^3 \times 0.1$ \cite{Chang2010}. These energy absorption and emission processes lead to decoherence when operating in the quantum regime, and for this reason it may be desirable to work in a cryogenic environment, or to work with internally cold particles.

Next we consider the cooling power due to collisions with gas molecules, again following \cite{Chang2010}:

% =================================================
\begin{equation}
\label{eqn:gascool}
\dot{E}_{\mathrm{gas}} = -\acccoeff \sqrt{\frac{2}{3\pi}}(\pi a^2)\vth \frac{\gamma_{\rm sh} + 1}{\gamma_{\rm sh} - 1}\left (\frac{\T{int}}{\T{gas}}-1 \right) \Pg,
\end{equation}
% =================================================

where $\vth$ is the mean thermal velocity of the impinging gas molecules, $\gamma_{\rm sh} = 7/5$ is the specific heat ratio of a diatomic gas, and for most experiments $\T{gas} \equiv \T{env}$. This expression holds in the Knudsen regime ($\bar{l}\gg a$). Combining all of these leads to a rate equation that describes $\T{int}$:

% =================================================
\begin{equation}
\label{eqn:intemp}
m\hcap \frac{\d\T{int}}{\d t} = \Iopt\abscross + \dot{E}_{\mathrm{gas}} + \dot{E}_{\mathrm{abs}}^{\mathrm{bb}} + \dot{E}_{\mathrm{emis}}^{\mathrm{bb}},
\end{equation}
% =================================================

where $\hcap$ is the specific heat capacity for the particle material and $\Iopt$ is the light intensity. Using eqn.~\eqref{eqn:intemp}, one can calculate the steady-state temperature of a sphere levitated in vacuum. It is also the case that the refractive index $n(\omega)$, and hence the permittivity $\eom{}$, of the levitated particle varies with the bulk temperature, but since for silica it varies by only $1\%$ over 2000\,K, we ignore this effect here. To avoid absorption, pairings of nanoparticle material and trapping wavelengths should be carefully chosen, for example working with pure silicon particles and telecoms wavelengths ($\sim 1500\,$nm).

\begin{figure}[b]
	\begin{center}
	{\includegraphics{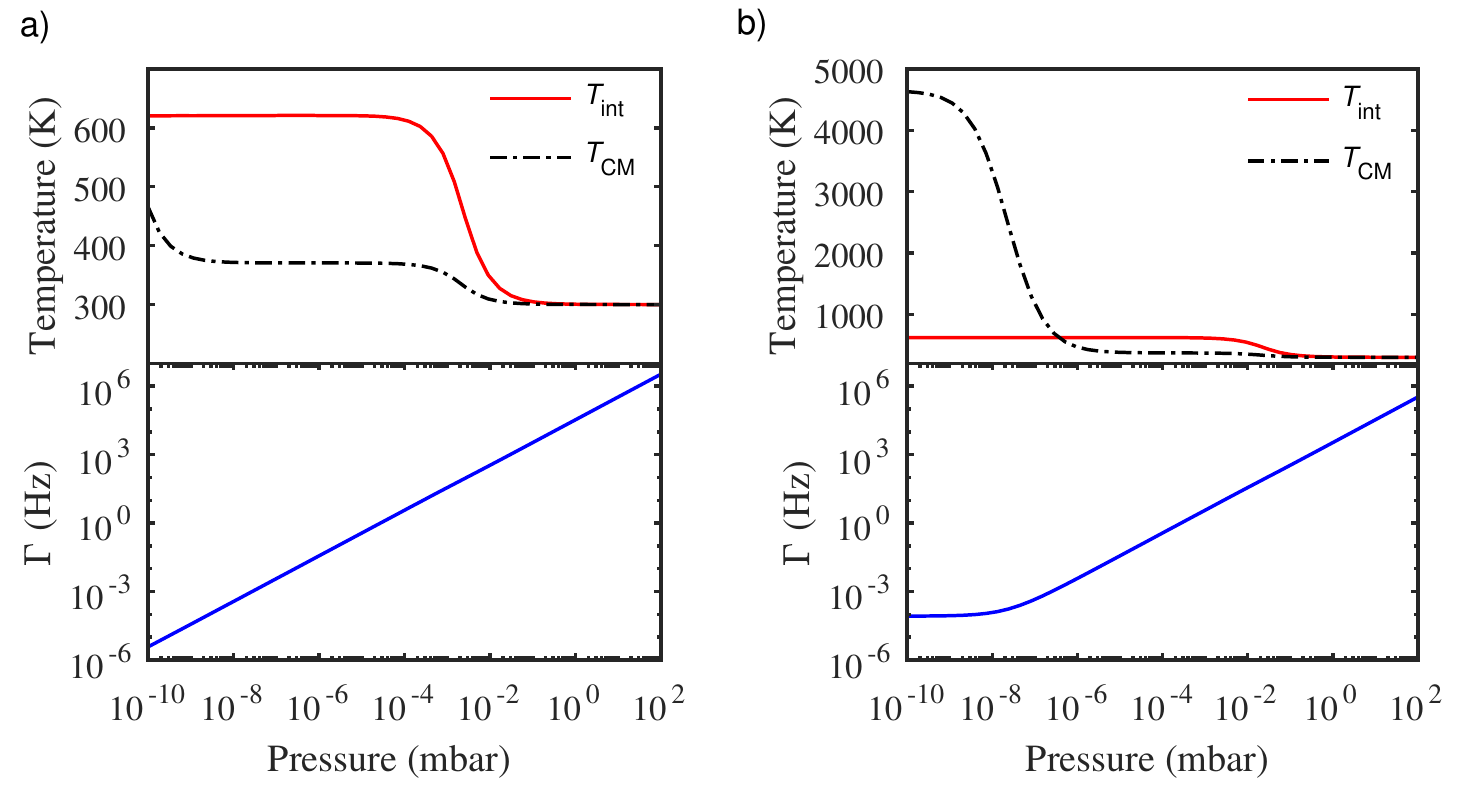}}
\caption{\label{fig:pressure} \textbf{The variation in particle dynamics with pressure.} Variation in the surface temperature $\T{int}$, centre-of-mass temperature $\T{CM}$, and damping rate $\g{CM}$ with pressure for a) $a = 10\,$nm, and b) $a = 100\,$nm silica spheres, with $\T{gas} \equiv \T{env} = 300\,$K. These dynamics are due to the balance between optical absorption, blackbody absorption and emission (eqn.~\eqref{eqn:bbenergy}), photon recoil heating (eqn.~\eqref{eq:radiation_damping}), and cooling due to collisions with gas molecules (eqn.~\eqref{eqn:gascool}). This figure assumes a sphere trapped with a realistic laser intensity of $6\times10^{11}\,$W\,m$^{-2}$ with a wavelength of 1550nm. The optical trap depth is a) $U_0/\kB = 520\,$K and b) $U_0/\kB = 5\times10^5\,$K. For silica we use a complex refractive index $n = 1.45 +(2.5\times10^{-9})i$ \cite{Bateman2014}, material density $\rho = 2198\,$kg\,m$^{-3}$, and we assume the surrounding gas is $N_2$, with a corresponding surface accommodation coefficient $\acccoeff = 0.65$.}
\end{center}
\end{figure}

In fig.~\ref{fig:pressure}, the variation in $\T{int}$ and $\T{CM}$ with pressure is shown, for silica spheres trapped in an optical tweezer under realistic experimental conditions. We can identify three regimes: At high pressures, the cooling power of the surrounding gas is sufficient to counter any heating due to optical absorption, and the particle surface and centre-of-mass temperatures thermalize to the environmental temperature. At low pressures, the surface temperature increases, due to reduced gas cooling power (eqn.~\eqref{eqn:gascool}) and at ultra low pressure, the centre-of-mass motion thermalizes with the photon shot noise \cite{Jain2016a}. 

It should be noted that all centre-of-mass heating mechanisms pose a problem to operating in the quantum regime. Because of this, most proposals for testing quantum physics with optically trapped particles involve switching off the light fields after state-preparation, and letting the particle drop. This may not be desirable for thermodynamics experiments which require long interrogation times. For this reason, some proposals consider magnetic or electric levitation, and even operation in space.

%%%%%%%%%%%%%%%%%%
%%%%%%%%%%%%%%%%%%%

\section{Brownian motion}
\label{sec:Brownian}

Besides its important role in the development of the foundations of physics, today the Brownian particle serves as an exemplary model to describe a variety of stochastic processes in many fields, including physics, finance and biology. Brownian motion in non-equilibrium systems is of particular interest because it is directly related to the transport of molecules and cells in biological systems. Important examples include Brownian motors, active Brownian motion of self-propelled particles, hot Brownian motion, and Brownian motion in shear flows.
Despite its importance, the first experimental observation of ballistic Brownian motion had to wait a century until Li \emph{et al.}'s seminal work with optically levitated microparticles \cite{Li2010}.
This result already highlights the importance of the levitated particle system for studying thermodynamics.\\
%This statement held until 2010, when Li \emph{et al.} \cite{Li2010} were able to observe for the first time the transition from diffusive to ballistic Brownian dynamics as shown in fig.~\ref{fig:BrownianMotion}.
%%They did this by monitoring the motion of a microparticle levitated in a gas at various pressures, where the particle-bath collision timescales ($1/\Gamma$) are much slower than for a particle in liquid.
%This result shows the importance of the levitated particle system for studying thermodynamics.

In this section we will discuss the basics of Brownian motion. We will mainly treat the aspects that are necessary for understanding the following discussion of thermodynamics with levitated nanoparticles. For details on the theory of Brownian motion we refer the reader to the work of Ornstein, Uhlenbeck and Wang \cite{Uhlenbeck:1930tn, Wang:1945wd}.
First we will consider the motion of a free Brownian particle. Then we will add the confining potential and discuss important concepts such as a the power spectral density. Then we will include higher order (nonlinear) terms of the trapping potentials and discuss how they impact the power spectra.
Finally, we go to the opposite extreme case where the particle is cooled to extremely low energies such that quantum effects have to be included.

\subsection{Free Brownian motion}
The three motional degrees of freedom of a free particle are decoupled and without loss of generality it suffices to discuss a single coordinate $q(t)$ ($q = x,y, z$).
When coupled to a thermal bath at temperature $\T{CM}$ with rate $\g{CM}$, the equation of motion is given by the Langevin equation

% ----------------------------------------------------------------------------------
\begin{equation}\label{eq:FreeLangevin}
\ddot{q} \,+\, \g{CM} \!\; \dot{q} =\; \mathcal{F}_q(t)/m, 
\end{equation}
% ----------------------------------------------------------------------------------
where  $\mathcal{F}_q(t)/m =\sqrt{2\kB \T{CM}\g{CM}/m}\,\whn(t)
$ and $\whn(t)$ is a normalised white-noise process with $\langle\whn(t)\rangle = 0$, $\langle\whn(t)\whn'(t)\rangle = \delta(t-t')$. Here $\delta(t-t')$ is the Dirac delta function.
Since $\mathcal{F}_q(t)$ is a random process, $q(t)$ is also a random variable, such that each trajectory starting from the same initial conditions is different. However, the mean and variance for an ensemble of particles are well defined and are identical  to the values for a single particle measured over a long time by virtue of the ergodic theorem.

Since the average force is zero, the mean particle position is also zero  $\langle q(t)\rangle = 0$.
Its variance, or mean-square displacement, is given by

% =================================================
\begin{equation}\label{eq:pos_variance_free}
\var{q}(t)=  \langle \left[q(t)- q(0)\right]^2\rangle = \frac{2\kB \T{CM}}{m\g{CM}^2}\left[\g{CM} t-1+e^{-\g{CM} t}\right].
\end{equation}
% eq. 10 from Uhlenbeck1930
% =================================================

At long time scales ($t\gg1/\g{CM}$), the variance is the same as that predicted by Einstein's theory of diffusion $ \var{q}(t) = 2 D t $ where $D = \kB \T{CM}\left/m\g{CM}\right.$ is the diffusion coefficient. This regime is truly random in the sense that the particle trajectory is fractal and, therefore, is continuous but not differentiable. At short time scales ($t \ll 1/\g{CM}$), the dynamics of a Brownian particle is dominated by its inertia and its trajectory is ballistic. In this regime, the variance grows quadratically in time $ \var{q}(t) = (\kB\T{CM}/m) t^2 $ as expected for a free particle.

\subsection{Harmonic Brownian motion}

Under the influence of trapping forces, the particle will be localised about its equilibrium position. For small displacements, the trap can be approximated by a three-dimensional harmonic potential. As before, we limit our discussion to a single coordinate. The equation of motion for a harmonically trapped Brownian particle is \cite{Wang:1945wd}

% ----------------------------------------------------------------------------------
\begin{equation}\label{eq:HarmonicLangevin}
\ddot{q} \,+\, \g{CM} \!\; \dot{q} +\wo^2 q=\;   \sqrt{2\kB \T{CM}\, \g{CM}/m}\,\whn(t). 
\end{equation}
% ----------------------------------------------------------------------------------

Due to the confinement provided by the trap the variance does not grow unbounded. Instead, the particle oscillates in the trap at the characteristic frequency $\tilde{\w} = \sqrt{\wo^2-\g{CM}^2/4}$. For the optical potential eqn.~\eqref{eq:potential}, the trap frequency $\wo$ is given by eqn.~\eqref{eqn:frequencies_translational}.

We distinguish between three cases, the overdamped ($\wo \ll \g{CM}$), the critically damped ($\wo \approx \g{CM}$) and underdamped case ($\wo \gg \g{CM}$). This stochastic equation of motion has been studied in detail by Ornstein and Uhlenbeck \cite{Uhlenbeck:1930tn} and we summarise their results here. The variance of the position of a Brownian particle in an under-damped harmonic trap is

% =================================================
\begin{equation}\label{eq:pos_variance_underdamped}
\var{q}(t)=  \frac{2\kB \T{CM}}{m\wo^2}\left[1-e^{-\frac{1}{2}\g{CM} t}\left(\cos(\tilde{\w} t) +\frac{\g{CM}}{2\tilde{\w}}\sin(\tilde{\w} t)\right)\right].
\end{equation}
% ========

In the over-damped harmonic trap, set $\tilde{\w}\to i \tilde{\w}$. In a critically damped harmonic trap, set $\tilde{\w}\to 0$. The position autocorrelation function is related to the variance as follows

% =================================================
\begin{equation}\label{eq:autocorrelation}
%\acorr{qq}(t)
%\langle q(t)q(0)\rangle
\corr{q}{q}
   = \frac{\kB \T{CM}}{m\wo^2} - \frac{1}{2}\var{q}(t)
\end{equation}
% =================================================

\begin{figure}[hbt]
  \includegraphics[width=0.8\textwidth]{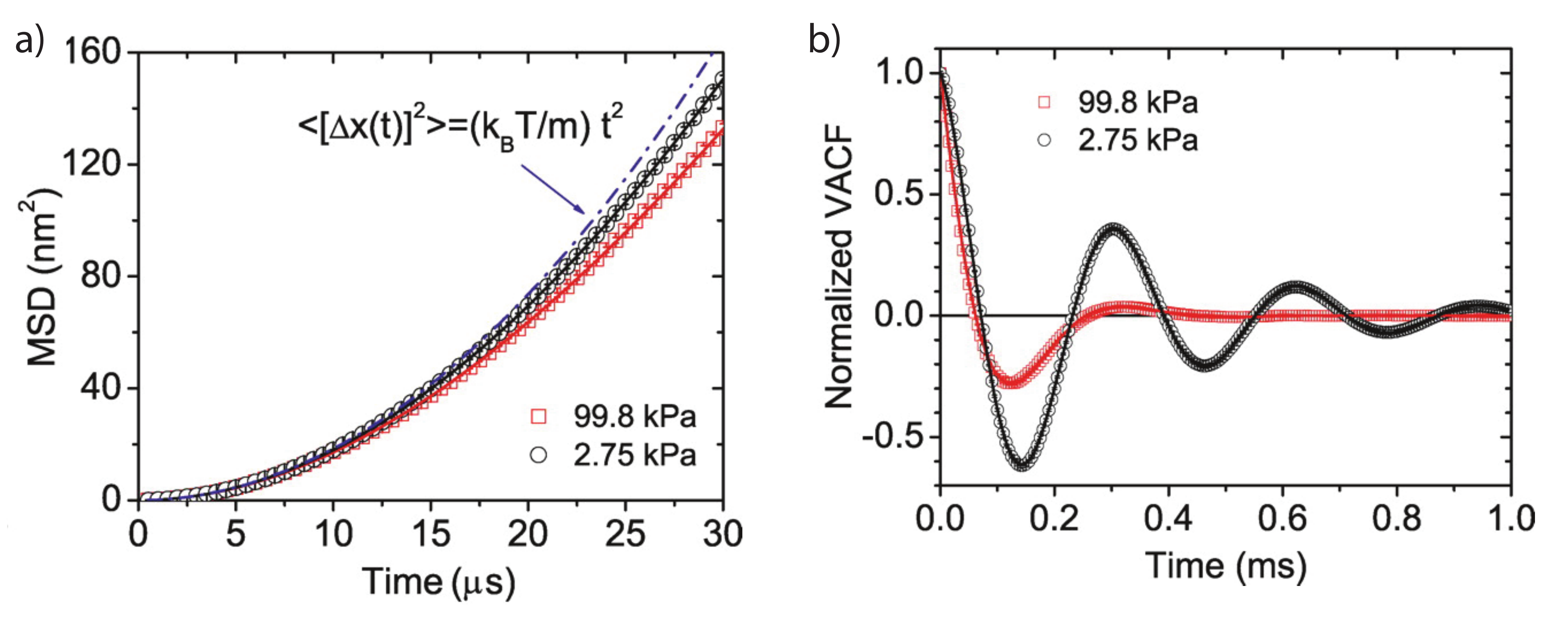}
  \caption{\textbf{First experimental observation of the instantaneous velocity of a Brownian particle.}
  a) The mean-square displacement for short times is proportional to $t^2$, a signature of ballistic motion.
  b) The normalised velocity autocorrelation functions for different pressures in perfect agreement with eqn.~\eqref{eq:vel_corr}. 
  Figures taken from \cite{Li2010} with permission from Science.
   \label{fig:BrownianMotion}}
\end{figure}

The velocity autocorrelation function is given by 

% =================================================
\begin{equation}\label{eq:vel_corr}
\corr{v}{v}=  \frac{\kB \T{CM}}{m}e^{-\frac{1}{2}\g{CM} t}\left(\cos(\tilde{\w} t) -\frac{\g{CM}}{2\tilde{\w}}\sin(\tilde{\w} t)\right),
\end{equation}
% see eq. 50c Wang1945
% ========

and an experimental verification of this form is shown in fig.~\ref{fig:BrownianMotion}. In addition, position and velocity are correlated and the position-velocity correlation function is given by

% =================================================
\begin{equation}\label{eq:velpos_corr}
\corr{q}{v}=\corr{v}{q}=  \frac{\kB \T{CM}}{m \tilde{\w}}e^{-\frac{1}{2}\g{CM} t}\sin(\tilde{\w} t).
\end{equation}
% see eq. 50b Wang194% ========

%\subsubsection{Power spectral density}
According to the Wiener-Khinchin theorem, the position autocorrelation function is the Fourier transform of the power spectral density $S_{qq}(\w) = \int_{-\infty}^\infty \corr{q}{q}e^{i\w t}\d t$, which for eqn.~\eqref{eq:HarmonicLangevin} is given by 

% =================================================
\begin{equation}\label{eq:PSD_position}
  S_{qq}(\w) = |\chi(\w)|^2 \Sff{}(\w)
  = \frac{\g{CM} \kB \T{CM}\left/\pi m\right.}{(\w^2-\wo^2)^2+\g{CM}^2\w^2}
\end{equation}
% =================================================

where 
$
  \chi(\w) = m^{-1}\left[\w^2-\wo^2+i\g{CM}\w\right]^{-1}
$
is the response function or susceptibility of a harmonic oscillator. In the underdamped regime, the frequency spectrum of the autocorrelation function is strongly peaked around the trap frequency $\wo$, where as when overdamped the frequency spectrum is broad, as shown in fig.~\ref{fig:PSD}a). The power spectral density (PSD) is a useful tool in experiments with harmonic oscillators, since the dynamics of the oscillator can be separated from (spectrally distant) noise. An analysis of the power spectral density allows one to extract the center of mass temperature of the oscillator and the damping rate, as is clear from eqn.~\eqref{eq:PSD_position}.

\subsection{Nonlinear Brownian motion}
\label{sec:non-lin-BM}
Until now, we have only considered small deviations from the equilibrium position, where the potential is harmonic. However, the actual trapping potential is nonlinear. For the transverse directions, the lowest order nonlinear term is a cubic or Duffing nonlinearity in the equation of motion due to the symmetry of the trap.
Along the direction of propagation of the trapping laser, the scattering force breaks the symmetry and we also get a quadratic term. Similarly, gravity breaks the symmetry along the $y$-axis (see fig.~\ref{fig:Schematic}).
However, due to the smallness of the quadratic nonlinearity we will neglect it and focus our discussion on the Duffing term. Including the latter, the equation of motion for a single coordinate reads

% ----------------------------------------------------------------------------------
\begin{equation}\label{eq:DuffingLangevin}
\ddot{q}_i \,+\, \g{i} \!\; \dot{q}_i +\w_i^2 q_i + \w_i^2\left(\sum_j\xi_{ij} q_j^2\right) q_i=\;   \sqrt{2\kB \T{CM}\,\g{i}/m}\,\whn(t), 
\end{equation}
% ----------------------------------------------------------------------------------

where we have re-introduced the indices for a clearer notation. From the optical potential eqn.~\eqref{eq:potential} we find that $\xi_{ij} \sim -\waist_j^{-2}$. As a consequence, the oscillation frequency becomes a function of the oscillation amplitude and is red shifted by \cite{Gieseler2013}

% =================================================
\begin{equation}\label{eq:freq_shift}
  \Delta\w_{\rm i} =\frac{3}{8}\w_i \sum_j \xi_{ij}  \a_{\rm j}^2,
\end{equation}
% =================================================

where $\a_i$ is the instantaneous oscillation amplitude. In the low damping regime ($\w\gg\g{CM}$), the amplitude $\a_i$ and phase $\phi_i$ are quasi-static over many oscillation periods $2\pi/\w$ and only change significantly over times scales on the order of the relaxation time $2\pi/\g{CM}$.
Hence, the position can be written as $q_i(t) = \a_i(\tau)\cos \left[\w t + \phi_i(\tau)\right]$, with $2\pi/\w \ll \tau\ll 2\pi/\g{CM}$, where $\tau$ represents the slow timescale of the amplitude and phase evolution.
The frequency shift due to changes in the oscillation amplitudes is also known as self-phase modulation ($j=i$) and cross-phase modulation ($j\neq i$). To resolve the nonlinear frequency shift originating from thermal motion, the nonlinear contribution must be larger than the linear one, resulting in the condition

% =================================================
\begin{equation}\label{eq:R_nonlinear_shift}
  \mathcal{R} = \frac{\Delta \w_{\rm NL}}{\g{CM}} = \frac{3\xi Q \kB \T{CM}}{4\w^2 m }\gg 1,
\end{equation}
% =================================================

where $Q = \w/\g{cm}$ is the quality factor. If this condition is fulfilled, the power spectral density (PSD) is no longer given by eqn.~\eqref{eq:PSD_position}. Instead, the harmonic oscillator PSD is now weighted with the probability to find the particle with a certain energy $E$ and the resulting PSD 

% =================================================
\begin{equation}\label{eq:PSD_NL}
  S_{\rm NL}(\w) =\int_0^\infty\rho(E)S_L(\w, E)\d E,
\end{equation}
% =================================================

is no longer symmetric, as shown in fig.~\ref{fig:PSD}b). The energy distribution is given by the Gibbs distribution $\rho(E) = Z^{-1}\exp(-E/\kB \T{CM})$ with $Z = \int \rho(E) \d E = \kB \T{CM}$
and the spectra $S_L(\w, E) = E \g{CM} \left/\pi m\wo^2\right.[(\w^2-\hat{\w}(E)^2)^2+\g{CM}^2\w^2]^{-1}$ are shifted to $\hat{\w}_0(E) = \wo+3\xi / (4 m\wo) E$.
Notably, due to the Gibbs distribution weighting term, the symmetry of the thermally driven spectra is opposite to the frequency response of the driven Duffing oscillator.

\subsection{Quantum Brownian motion}
\begin{figure}[hbt]
  \includegraphics[width=\textwidth]{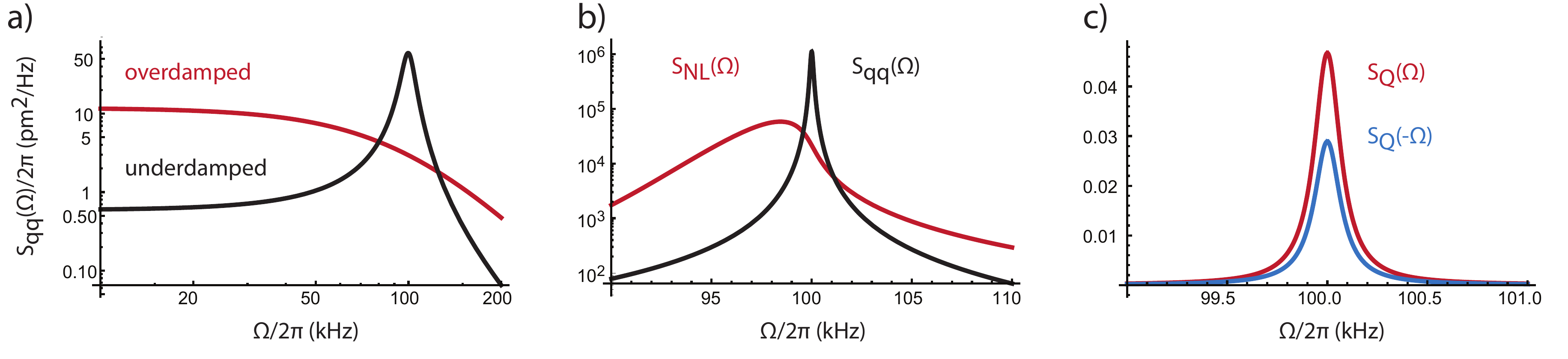}
  \caption{\textbf{Position power spectral densities (PSD) in different regimes.}
  a.) In the overdamped regime (quality factor $Q = \w/\g{cm} = 1/2$, red) the PSD has its maximum at $\w=0$ and falls off for higher frequencies. In the underdamped regime ($Q$ = 10, black), the PSD is peaked around the resonance frequency $\wo$ and has a linewidth of $\g{} \approx \wo/Q$.
  b) For even higher $Q$, nonlinear effects can broaden the linewidth and instead of the expected narrow harmonic oscillator PSD (black, $Q$ = 100, eqn.~\eqref{eq:PSD_position}), the observed PSD is highly asymmetric (red, $\xi = 5\,\rm \mu m^{-2}$, $T=300\,\rm K$,  eqn.~\eqref{eq:PSD_NL})
  c) When the motion is cooled near the quantum ground state (here $\T{CM} = 10\,\rm \mu$K), nonlinear effects are negligible. Instead, quantum features lead to an asymmetric PSD, where the PSD at positive frequencies (red) is by a factor $\exp(\hbar \wo / \kB\T{CM}.)$ higher than the PSD at negative frequencies (blue).
   \label{fig:PSD}}
\end{figure}

While the nonlinear aspects of the potential are only relevant for large excitations, the opposite extreme, when the center of mass temperature is of the order of a single quantum of motion $\kB \T{CM}\approx \hbar \w_0$, is of particular importance since quantum effects can no longer be neglected.
In the quantum regime, the position autocorrelation eqn.~\eqref{eq:autocorrelation} contains the product of time-evolved operators $\langle \hat{q}(t) \hat{q}(0)\rangle$, which do not commute.
As a result, the spectrum

% =================================================
\begin{equation}
	\label{eq:PSD_quantum}
  S_{Q}(\w) %
  = \frac{\hbar/\pi}{1-\exp\left(-\frac{\hbar \w}{\kB \T{CM}}\right)} {\rm Im} \chi(\w)
= \frac{\hbar\w m\g{eff}/\pi}{1-\exp\left(-\frac{\hbar \w}{\kB \T{eff}}\right)} |\chi(\w)|^2
\end{equation}
% =================================================

is asymmetric in frequency, where the PSD at positive frequencies is a factor $\exp(\hbar \wo\left/\kB \T{CM} \right.)$ higher than the PSD at negative frequencies, as shown in fig.~\ref{fig:PSD}c). The positive-frequency part of the spectral density is a measure of the ability of the oscillator to absorb energy, while the negative-frequency part is a measure of the ability of the oscillator to emit energy.
Therefore, we can understand the positive frequency part of the  spectral density as being related to stimulated emission of energy into the oscillator, while the negative-frequency part is related to the emission of energy by the oscillator.

Typically, the motional frequencies of a levitated particle are $\sim 100\,\rm kHz$. Therefore, the required temperature is a few micro-kelvin and therefore out of reach for cryogenic techniques and one has to resort to active cooling techniques. Recent experiments using feedback cooling have already attained motional occupations of a few tens of phonons \cite{Jain2016a}.

\section{Time dependent potentials}
\label{sec:time_dep}
So far we have considered only static trapping potentials $U_{\rm opt} \equiv U_{\rm opt}(\mathbf{r})$, where the trapping laser power is constant. However, through modulation of the trapping beam intensity, we can make the optical potential time-dependent. This is particularly useful when studying non-equilibrium dynamics. From eqn.~\eqref{eqn:frequencies_translational} it follows that a change in optical power $\delta \Popt(t)$ changes the trap frequency by $\w = \wo(1+\emod(t)/2)$, where $\emod(t) = \delta \Popt(t) \left/ \Poptm\right.$ and $\Poptm$ is the mean optical power. The equation of motion under this parametric modulation is given by 

% ----------------------------------------------------------------------------------
\begin{equation}\label{eq:ParametricLangevin}
\ddot{q} \,+\, \g{CM} \!\; \dot{q} +\wo^2\left[1+\emod_0 \cos(\omegamod t)+\xi q^2 \right]q=\;   \sqrt{2\kB \T{CM}\g{CM}/m}\,\whn(t) .
\end{equation}
% ----------------------------------------------------------------------------------

 Energy is most effectively exchanged between the trapping laser and the particle if the modulation $\emod(t) = \emod_0\cos(\omegamod t)$ occurs at twice the trapping frequency $\omegamod\approx 2\w_0$. The flow of energy is thereby determined by the relative phase $\phimod$ between the particle oscillation and the laser intensity modulation\footnote{Note that $\phimod$ doesn't appear in the expression for $\emod(t)$, since the modulation serves as the time reference and $\phimod$ is the phase of the particle with respect to the modulation.}. If the modulation is in-phase, energy is extracted (cooling), while the motion is excited when the modulation is out-of-phase (heating). For $\emod_0>1/2Q$, where $Q = \omega_0/\g{cm}$ is the motional quality factor, energy is pumped into the system faster than can be extracted through dissipation. For a harmonic trap this would lead to a steady increase in energy. However, due to the nonlinear Duffing term, the oscillation frequency of the particle shifts away from the energy matching condition, which limits the maximum oscillation amplitude.
Without active stabilization of the modulation phase with respect to the particle motion, the relative phase is random. Therefore, to achieve cooling the phase needs to be actively stabilized, for instance with a phase-locked-loop \cite{Jain2016a}.

\subsection{Effective potentials and non-equilibrium steady states}
\label{sec:effective_potential}

As we discussed in Sec.~\ref{sec:non-lin-BM}, the particle motion is described by a slowly varying evolution of the phase and amplitude and a fast modulation at frequency $\w$. 
In many cases we are primarily interested in the slow evolution of the energy or amplitude and it is, thus, advantagous to work with the effective equations of motion for the energy instead of considering to full particle dynamics.
This strategy allows us to define effective potentials for the energy and to derive an effective temperature for the particle center-of-mass motion.

For convenience we introduce the position $q$ and momentum $p=m\dot q$ differential equations of motion for a particle in a time-dependent potential

% ----------------------------------------------------------------------------------
\begin{subequations}
\begin{align}
\d q&=\frac{p}{m}\d t, \label{equ:SDE1}\\
\d p &=\left[-m\wo^2q-\g{CM} p+\emod_0 m \Omega^2_0 \cos(\omegamod t)q\right]\d t+\sqrt{2m\g{CM} \kB \T{CM}}\,\d W.
\label{equ:SDE2}
\end{align}
\end{subequations}
% ----------------------------------------------------------------------------------
Here, $W(t)$ is the Wiener process with $\langle W(t) \rangle  =  0$, $\langle W(t) W(t') \rangle  =  t'-t$. Note that $\langle W^2(t) \rangle = t$ for any time $t\ge 0$ and, thus, for an infinitesimal time interval $\d t$ one has $\langle ({\rm dW)^2}\rangle=\d t$. The white noise $\whn(t)$ appearing in the random force can be viewed as the time derivative of the Wiener process, $\whn(t)= \d W(t)/\d t$. The total energy of the particle in one dimension is given by

% ----------------------------------------------------------------------------------
\begin{equation}\label{eq:Energy}
E(q,p)=\frac{1}{2}m\wo^2 q^2+ \frac{p^2}{2m}+\frac{1}{4}\xi m\wo^2 q^4.
\end{equation}
% ----------------------------------------------------------------------------------

To avoid multiplicative noise, i.e. a noise term that depends on the current value of the energy, we consider the square root of the energy rather than the energy itself,
% ----------------------------------------------------------------------------------
\begin{equation}
\sqrtE(q, p)=\sqrt{E(q, p)}.
\end{equation}
% ----------------------------------------------------------------------------------

At low friction, the amplitude $\a$ and phase $\phimod$ with respect to the driving force are quasi-constant, and the particle performs an undisturbed harmonic oscillation evolving according to

% ----------------------------------------------------------------------------------
\begin{equation}
q(t)=\a \cos (\Omega t + \phimod) \qquad \qquad p(t)=-m\Omega \a\sin(\Omega t + \phimod),
\end{equation}
% ----------------------------------------------------------------------------------

where the amplitude of the oscillation is related to $\sqrtE$ by $\a=\sqrt{2/m}(\sqrtE/\Omega)$. Note that the oscillation frequency $\w$ is not necessarily the same as the frequency $\wo$ of the unperturbed harmonic oscillator. For instance, for strong modulation the particle motion entrains with the modulation and $\w \approx \omegamod/2$ \cite{Gieseler:2014wt}.

Applying Ito's formula for the change of variables to $\sqrtE(q, p)$ and integrating over an oscillation period, we find that the change $\d \sqrtE$ during a short time interval is given by a Langevin equation for a \emph{fictitious} overdamped Brownian particle 

% ----------------------------------------------------------------------------------
\begin{equation}\label{eq:stochastic_epsilon}
\d   \sqrtE = \geff^{-1}f(\sqrtE)\d t+ \sqrt{2 \kB \T{CM}\left/\geff\right.} \d W
\end{equation}
% ----------------------------------------------------------------------------------

with damping $\geff = 4/\g{CM}$, moving through an effective potential

% ----------------------------------------------------------------------------------
\begin{equation}\label{eq:potential_epsilon}
\Ueff(\sqrtE)=\sqrtE^2 
-\kB \T{CM} \ln \sqrtE
+\frac{\sqrtE^2\emod_0\wo^2 \sin(2\phimod)}{2\g{CM}\w },
\end{equation}
% ----------------------------------------------------------------------------------

under the influence of an external force $f(\sqrtE) = - \d \Ueff(\sqrtE)/\d \sqrtE$. Note that due to the integration over one oscillation period, this equation has $\sqrtE$ as its only time dependent variable, while the dependence on other variables has been removed. By virtue of this isomorphism with over-damped Brownian motion, one can then immediately infer that eqn.~\eqref{eq:stochastic_epsilon} samples the distribution 
$
P_\sqrtE(\sqrtE) \propto \exp\left\{-\beta \Ueff(\sqrtE)\right\},
$
where $\beta = 1\left/\kB \T{CM}\right.$. Equation~\eqref{eq:stochastic_epsilon} implies that the time evolution of $\sqrtE$ can be viewed as a Brownian motion in the high friction limit.
A small real friction $\g{CM}$ corresponds to large effective friction $\geff$ determining the time evolution of $\sqrtE$ and, thus, the energy $E$ of the oscillator.
Interestingly, the \emph{fictitious} Brownian particle of the \emph{time-dependent} optical potential can exhibit similar dynamics to the \emph{real} Brownian particle in a \emph{static} optical potential \cite{Ricci:2017eh}.

Changing variables from $\sqrtE$ to $E=\sqrtE^2$ and applying Ito's formula, we finally obtain the probability density function of the energy, 
$
P_E(E) = \frac{1}{Z} \exp\left\{-\beta' E\right\}
$,
where $\beta' = 1/\kB\T{CM}'$ with effective temperature 

% =================================================
\begin{equation}\label{eq:Teff}
%\boxed{ 
\T{CM}' = \T{CM}\left(1+\frac{\emod_0\wo^2 \sin(2\phimod)}{2\g{CM}\w}\right)^{-1}.
%}.
\end{equation}
% =================================================

Equation~\eqref{eq:Teff} states that parametric modulation of the trapping potential results in an effective temperature change of the environment, where the particle centre-of-mass temperature changes from $\T{CM}$ to $\T{CM}'$. For $-\pi/2<\phimod<0$, $\T{CM}'>\T{CM}$, that is the particle motion is heated, while for $0<\phimod<\pi/2$, $\T{CM}'<\T{CM}$ and the particle motion is cooled. The rate at which the particle thermalizes with this effective bath is $\g{CM}'   = \g{CM} \left(\T{CM}/\T{CM}'-1\right)$, where the largest rates are achieved at $\phimod = -\pi/4$ and $\phimod = \pi/4$, for heating and cooling respectively. If the relative phase between the particle motion and the modulation $\phimod$ is not stabilized actively, for example through implementing a phase-locked loop fed back onto the trapping laser intensity, the particle motion will self-lock to $\phimod = -\pi/4$. Thus, a effective hot bath can be implemented easily by a simple modulation of the trapping laser at $\omegamod\approx 2\wo$.

The change of variables also yields the corresponding stochastic differential equation for the energy

% =================================================
\begin{equation}
{\rm d} E = \left[-\g{CM} (E - \kB \T{CM})
 -\frac{\eta  \wo E^2}{2m \w^2}
-\frac{E\emod\wo^2 \sin(2\phi)}{2\w}
\right]{\rm d}t 
+\sqrt{2E \g{CM} \kB \T{CM}}{\rm d}W. 
\label{equ:SDE_energy}
\end{equation}
% =================================================

In contrast to the stochastic equation of motion for $\sqrtE$, here the noise is multiplicative, i.e., its amplitude is energy dependent.

It is predicted that by engineering an effective cold bath, it is possible to cool the motion of a levitated nanoparticle to its motional quantum ground state $\kB\T{CM} \leq \hbar\w$, for operation in the quantum regime. One method is to use the passive feedback provided by optical cavity cooling, with firm predictions of reaching the quantum ground-state \cite{Chang2010}. The thermal occupation of an optical cavity at room temperature is extremely low, with a photon occupation of $n_{\rm ph} = \sqrt{\kB\T{env}/\hbar\omL}\ll 1$, which forms the effective low temperature bath. Using active feedback cooling $\sim 100\, \mu$K temperatures have been achieved, corresponding to a phonon occupancy of $\sim 20$ \cite{Jain2016a}.

\subsubsection{Non-thermal states}
\label{sec:non-thermal}
The modulation of the trapping potential gives rise to a non-conservative force that allows us to inject and extract energy from the particle. Since there is a continuous flow of energy, the particle is no longer in thermal equilibrium. Surprisingly, under the appropriate conditions we can describe the particle as in thermal equilibrium with an effective bath (c.f. eqn.~\eqref{eq:Teff}). However, this description breaks down when we heat the particle ($\phimod = -\pi/4$) above the threshold condition 
$
\emod > 2\Qf^{-1}\sqrt{1+\Qf^2\left(2-\omegamod/\wo \right)^2}\approx 2\Qf^{-1}
$, where the approximation is exact at parametric resonance $\omegamod = 2\wo$. Then the effective temperature diverges and the motion transitions from a thermal state to a coherent oscillation, which is phase-locked to the modulation source, similar to the lasing condition of an optical oscillator.

A more subtle non-equilibrium steady state, which can no longer be described an effective thermal bath can be achieved by parametric feedback modulation of the form $\emod_\text{fb}(t) = -(\fb/\wo) q(t)\dot{q}(t)$, where $\fb$ parametrizes the feedback strength. This leads to a parametric modulation at the parametric resonance condition, while ensuring a phase that is optimized for extracting energy from the mechanical mode. However, in contrast to the previous case with constant modulation amplitude, here the modulation amplitude is proportional to the particle energy $\emod_0 \propto \a^2\propto E$. As a consequence, the particle feels a nonlinear friction force with $\g{NL}\propto E$.

The probability distribution for the energy, including the position dependent feedback term $\fb$ and parametric modulation with constant amplitude $\emod_0$, is then given by

% =================================================
\begin{equation}
P_E(E) = \frac{1}{Z} \exp\left\{-\beta\left[\left(1+\frac{\emod_0\wo^2 \sin(2\phimod )}{2\g{CM}\w}\right)E
+\frac{\eta  \wo}{4m\g{CM} \w^2}E^2 \right]\right\},
\label{eq:energy_distribution}
\end{equation}
% =================================================

where the normalization factor $Z=\int P_E(E){\rm d}E$ is given by

% =================================================
\begin{equation}
Z = \sqrt{\frac{\pi m\g{CM} \w^2}{\beta \eta  \wo}}
 h \left(\sqrt{\frac{\beta m\g{CM} \w^2}{\eta  \wo}}\left(1+\frac{\emod_0\wo^2 \sin(2\phimod)}{2\g{CM}\w}\right) \right),
\end{equation}
% =================================================

and the function $h(x)$ is defined as
$
h(x)=\exp(x^2){\rm erfc}(x)
$
and ${\rm erfc}(x)$ is the complementary error function. Thus, the energy distribution is that of an equilibrium system with effective energy

% =================================================
\begin{equation}
%\boxed{
H=\left[1+\frac{\emod_0\wo^2 \sin(2\phimod)}{2\g{CM}\w}\right]E+\frac{\eta  \wo}{4m\g{CM} \w^2}E^2.
%}.
\end{equation}
% =================================================

While the term proportional to $E^2$ is caused by the feedback cooling, the term proportional to $E$ is affected only by the parametric modulation. 
Since, for low friction, the energy of the oscillator is essentially constant over many oscillation periods, the full phase-space density $P_{q, p}$ can be obtained by averaging the micro-canonical distribution $P_m(q,p;\tilde{E}) = g^{-1}(\tilde{E})\delta\left[E(q,p)-\tilde{E}\right]$ over the energy distribution eqn.~\eqref{eq:energy_distribution}. For low friction constants and small feedback strength, this linear superposition of micro-canonical distributions is valid even under non-equilibrium, conditions  and we obtain

% =================================================
\begin{equation}\label{eq:Position_distribution_SS}
  P_{q, p}(q, p) =\frac{\wo}{2\pi} P_E(E(q,p)),
\end{equation}
% =================================================

where the $E(q,p)$ is given by eqn.~\eqref{eq:Energy}, and we approximated the micro-canonical density of states with the density of states for the harmonic oscillator  $g(\tilde{E}) \approx 2\pi/\wo$, that is we neglect the Duffing term of the potential in  $g(\tilde{E})$. Note, however, that while we have neglected the Duffing term in the expression for the density of states, it is included in the energy appearing in the argument of the exponential on the right-hand side of the above equation~\eqref{eq:energy_distribution}.

\subsubsection{Thermal squeezing}
A big advantage when using a levitated oscillator over conventional nano-mechanical oscillators is that the mechanical frequency can be changed by changing the power of the trapping laser. This allows one to perform unconfined time-of-flight measurements, to create physically large superposition states \cite{Bateman2014}, and to prepare squeezed states. In the quantum regime, squeezing enables one to push the fundamental quantum uncertainty below the standard quantum limit. A thermal state can be squeezed to reduce the uncertainty in one of the quadratures at the expense of anti-squeezing the other. While classical thermal squeezing of a levitated particle has been observed experimentally \cite{Rashid:2016hp}, squeezing below the standard quantum limit remains elusive. In the following we discuss how a change in laser power leads to squeezing.

A sudden change, or quench, in power of the trapping beam changes the mechanical frequency to a new value $\w \to \ws$. Thus, the time evolution of the position and momentum of the harmonic oscillator is given by

%=================================================
\begin{subequations}
\begin{align}
  q(\tau \ws) &= X_Q\cos(\ws \tau)+P_Q\frac{\w}{\ws }\sin(\ws \tau)\\
  p(\tau \ws) &= m P_Q \w \cos(\ws \tau)-m X_Q\ws\sin(\ws \tau),
\end{align}
\end{subequations}
% =================================================

where we introduced the position and momentum quadratures $X_Q = q(0)$ and $P_Q = -\dot{q}(0)/\w$ to denote the position and velocity at the time of the quench. After a time $\tau$, the power is switched back to its original value and we find that the phase space distribution for position and momentum is 

% =================================================
\begin{align}\label{eq:Position_distribution_SQ}
  P_{q,p}^\text{sq}(q, p, \tau) &= \frac{\w \beta}{2\pi} \times \\
    &\exp\left[-\beta
   \frac{1}{2}m\Omega^2
   \left(
   \left[X_Q \cos(\ws\tau)+e^{2r}P_Q \sin(\ws\tau)\right]^2 
+
\left[P_Q  \cos(\ws\tau)-e^{-2r}X_Q \sin(\ws\tau)\right]^2 \right)
  \right]\nonumber
 \end{align}
% =================================================

where we introduced the squeezing parameter $r = \frac{1}{2}\log(\w/\ws)$. Therefore, the squeezing pulse of duration $\tau$ leads to a non-Gaussian state, with correlations between position and momentum.
However, at times $\tau = \pi/ 2\ws$, the exponent in eqn.~\eqref{eq:Position_distribution_SQ} simplifies and we find that the position quadrature is squeezed by a factor $\exp(-2r)$, while the momentum quadrature is anti-squeezed by $\exp(2r)$.
Due to the reduced width of the squeezed distribution along a particular direction, this kind of state preparation allows one to reduce the measurement uncertainty. However, to be actually useful, the error introduced by the anti-squeezing of the momentum quadrature should not overwhelm the squeezing of the position quadrature.
Note that in contrast to the distribution eqn.~\eqref{eq:Position_distribution_SS}, which describes a steady state distribution, i.e. it does not depend on the observation time, eqn.~\eqref{eq:Position_distribution_SQ} is defined at a specific time (right after the application of the squeezing pulse). From this distribution, the system will relax back into thermal equilibrium as described in section \ref{sec:Relaxation}.

\section{Thermodynamics}
\label{sec:thermodynamics}
In this final section we will discuss the application of levitated nanoparticles to problems in stochastic thermodynamics and highlight some relevant experimental results.
%First we consider a static potential but with at least to local minima instead of the single trapping site that we have discussed so far. This leads us to the Kramers turnover problem, which similar to the Brownian particle is fundamental to the description of stochastic processes in nature. 
%Then we will consider how the particle relaxes from non-equilibrium states and 

\subsection{Kramers escape and turnover}

We have discussed the dynamics of a particle confined within a potential, and subject to fluctuating forces from the environment. Due to the stochastic nature of the imparted force, there is a probability that the particle will gain enough energy to escape the potential, even when it is confined by a potential much deeper than $\kB\T{CM}$, in a process known as Kramers escape. This form of ``classical tunnelling'' appears in a diverse range of physical systems, importantly including chemical reaction rates, protein folding, atomic transport in optical lattices and molecular diffusion at solid-liquid interfaces \cite{Rondin2017}.

The Kramers' escape rate is given by
% =================================================
\begin{equation}\label{eq:KramersRate}
  \R{K} = \R{0} \exp\left(-\frac{U_{\rm opt}}{\kB \T{CM}}\right)
\end{equation}
% =================================================
where $\R{0}$ is the attempt frequency and $U_{\rm opt}$ is the barrier height. From the Boltzmann factor in eqn.~\eqref{eq:KramersRate} it follows that such a transition is exponentially suppressed if the potential is much deeper that the thermal energy $U_{\rm opt}\gg \kB \T{CM}$.
Closely related to the Kramers escape is the Kramers turnover problem, which describes the tunnelling between two potential minima as the friction is varied. This is often more relevant in physical situations, describing the transitions between two protein configurations, for example. Kramers found \cite{Kramers1940} that in the underdamped regime, the transition rate increases with \emph{increasing} friction, and that in the overdamped regime the transition rate increases with \emph{decreasing} friction, with the transition region labelled the turnover. Fifty years later, a theory was developed that linked the two regimes \cite{Melnikov1991}. The first quantitative measurement of Kramers turnover was achieved using a levitated nanoparticle hopping between two potential wells formed by focussed laser beams. In this experiment, the friction rate was varied over many orders of magnitude through a change in the gas pressure $\Pg$ \cite{Rondin2017}.

\begin{figure}[hbt]
	\includegraphics[width=0.8\textwidth]{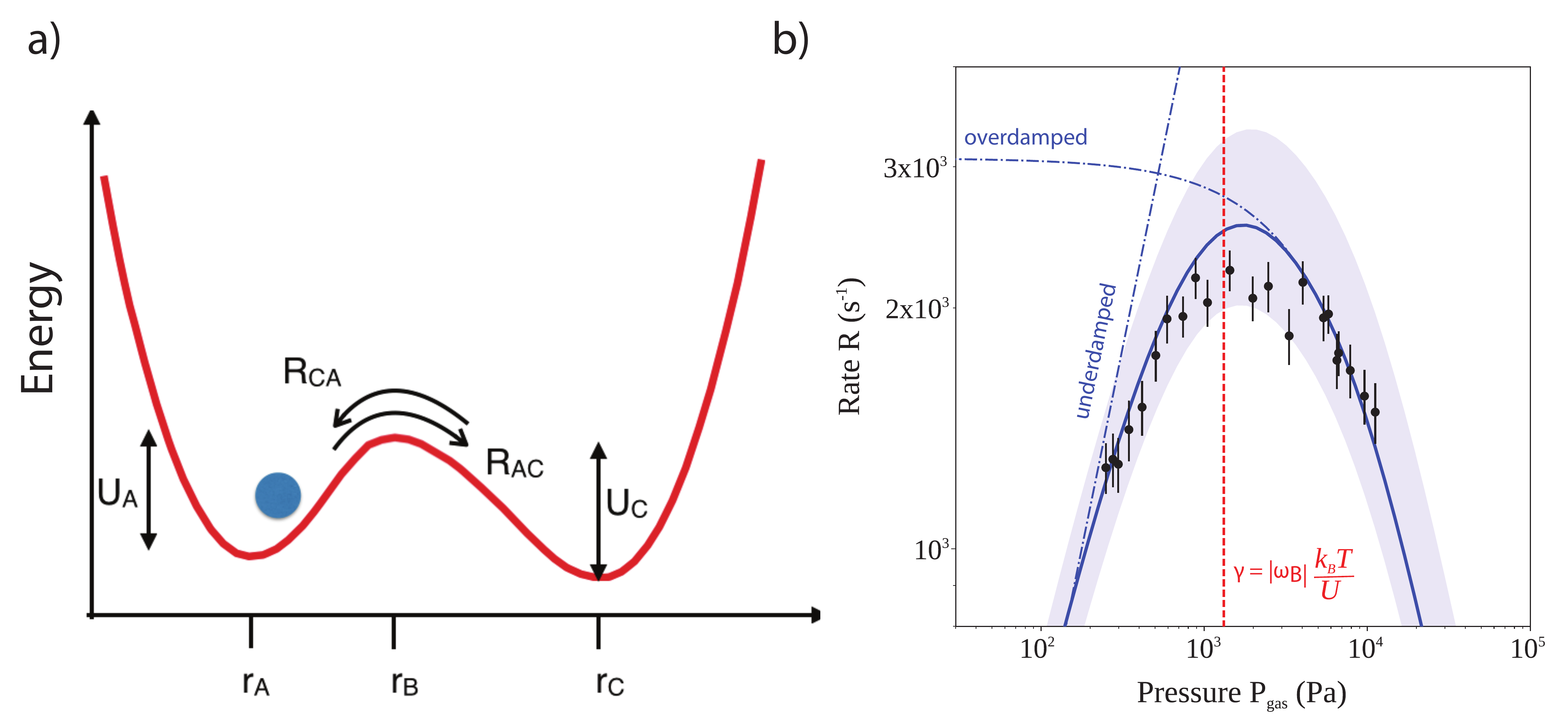}
  \caption{\textbf{Measurement of the Kramers turnover with a levitated nanoparticle.}
  a.) Illustration of a particle in a (generally asymmetric) bistable potential. The hopping rate $R$ between the wells $A, C$ depends upon the local potential $U$, and the background pressure $\Pg$.
  b.) Data illustrating the first experimental observation of Kramers' turnover, taken from \cite{Rondin2017}. Also marked is the full theory from \cite{Melnikov1991} (solid line), the turnover point which depends on the comparison of the damping rate $\g{CM}$ and the harmonic trap frequency at the crossing point $\Omega_{\rm B}$ (dashed line), and the limiting cases as predicted by Kramers \cite{Kramers1940} (dot-dashed lines). 
   \label{fig:Kramers}}
\end{figure}

We consider the hopping between two metastable potential wells that are separated by a barrier. The local principal axes are labeled $i=x',y',z'$ and the potential extrema are labeled $p = A,B,C$, as illustrated in fig.~\ref{fig:Kramers}a). Since the particle is not lost but recaptured in the second well, this problem is much more convenient to study experimentally than stochastic escape. A double well potential can be created by using two tightly focused laser beams, where the intensity and exact relative position of the two foci determines the height of the barrier. The hopping rates between the two wells is determined by the local curvatures of the potential at the extrema.

In the overdamped regime, the hopping from well $A$ to well $C$ via the barrier $B$ is given by Kramers' law. For a three-dimensional optical potential its dependency on the potential parameters is given by 
 
% =================================================
\begin{equation}\label{eq:R_AC_HD}
    R_{A\to C}^{\rm HD} = \frac{1}{2\pi} \prod_{i\in\{x',y',z'\}}  \frac{\w_i^A}{|\w_i^B|} \left[ \sqrt{|\w_B^S|^2+\frac{\g{CM}^2}{4}}-\frac{\g{CM}}{2}\right] e^{-\frac{U^A}{\kB\T{CM}}} 
    \approx \frac{1}{2\pi} \frac{\w^A\w^B}{\g{CM}} e^{-\frac{U^A}{\kB\T{CM}}} \ ,
\end{equation}
% =================================================

where $\w^p_i$ are the three frequencies at the three extrema along the local principal axis ($x',y',z'$) and $\w_B^S$ is the purely imaginary frequency of the saddle point \cite{Rondin2017}. The approximation holds in the limit of high damping $\g{CM}\gg\w_B$ and one dimensional motion.

In the underdamped regime, on the other hand, the rate is limited by the slow transfer of energy between the particle and its environment. This leads to a hopping rate that is proportional to $\g{CM}$

% =================================================
\begin{equation}\label{eq:R_AC_LD}
    R^{\rm LD}_{A\to C}=\frac{\g{CM} S^A}{2\pi}\frac{\w^A}{\kB\T{CM}} e^{-\frac{U^A}{\kB\T{CM}}}.
\end{equation}
% =================================================

where $S^p=4\int_{r_p}^{r_B}\sqrt{2m(U_B-U(\mathbf{r}))} \d r$ is the particle action over one oscillation period in well $p$ and is measured along the minimum energy path of the potential. These two limiting cases were already derived by Kramers \cite{Kramers1940}. In the transition region, such a simple analytical formula does not exist. Instead the rates depend on the depopulation factor

% =================================================
\begin{equation}\label{eq:depopulation_factor}
    \Upsilon(\delta) =  \exp \left[\frac{1}{\pi}  \int_{0}^{\infty}\ln\left\{ 1-\exp\left[-\frac{\delta}{k_B T} \left(x^2 +\frac{1}{4}\right) \right] \right\}\frac{\mathrm d x}{x^2 +\frac{1}{4}} \right]  \ ,
\end{equation}
% =================================================

where $\delta$ is the energy loss parameter. Generally, the estimation of the energy loss parameter is quite challenging. However, for memory-free friction, the energy loss is well approximated by $\delta = \g{CM} S_p$.

To account for the difference in transition rates from well A to C versus well C to A we need to multiply the transition rates by a factor $\prod_{p=A,B}\Upsilon(\g{CM} S^p)/\sum_{p=A,B}\Upsilon(\g{CM} S^p)$ and arrive at the general expression for the hopping rate 

% =================================================
\begin{equation}\label{eq:R_hopping}
    R (\g{CM}) = \frac{\Upsilon(\g{CM} S^A) \Upsilon(\g{CM} S^C)}{\Upsilon(\g{CM} S^A+\g{CM} S^C)}\left[ R_{A\to C}^{\rm HD}+ R_{C\to A}^{\rm HD}\right]\, .
\end{equation}
% =================================================

Figure~\ref{fig:Kramers}b) shows the limiting cases in the high and low damping regime, and the full solution for arbitrary damping. In addition, the figure includes experimental data from Rondin \emph{et al.} which, using an optically levitated nanoparticle, presents the first quantitative measurement of the Kramers rate across the turnover \cite{Rondin2017}.

\subsection{Relaxation \label{sec:Relaxation}}
In the steady-state, a trapped particle samples the distribution eqn.~\eqref{eq:energy_distribution}, which depends on experimental parameters, such as the average power of the trapping laser, and the rate and depth of any modulation of the optical potential. Hence, under a non-adiabatic change of the parameters, the systems relaxes into a new steady state.

The Fokker-Planck equation that describes the time evolution of the probability density function $P_E(E, t)$, including feedback and modulation, is given by

% =================================================
\begin{equation}\label{eq:FokkerPlanckEnergy}
\frac{\partial P_E(E, t)}{\partial t}=\frac{\partial }{\partial E}\left[\g{CM} (E - \kB T)
 +\frac{\eta  \wo E^2}{2m \Omega^2}
 +\frac{E\emod_0\wo^2 \sin(2\phimod)}{2\w}\right]P_E(E, t)
 +{\g{CM} \kB T}\frac{\partial^2 }{\partial E^2}E P_E(E, t).
\end{equation}
% =================================================

In general it is non-trivial to find an analytic solution to eqn.~\eqref{eq:FokkerPlanckEnergy}.
Amazingly, in the absence of feedback cooling ($\fb = 0$), the equation of motion for the energy corresponds to the 
Cox-Ingersoll-Ross model for interest rates, for which the exact analytical solution is given by the Noncentral Chi-squared distribution \cite{Salazar:2016ey}

% =================================================
\begin{equation}
  P_E(E|E_0, t) = c_t e^{-c_t(E+E_0e^{-\g{CM} t})}I_0(2c_t\sqrt{E E_0 e^{-\g{CM} t}}),
  \end{equation}
% =================================================

where $c_t = \beta\left(1-e^{-\g{CM} t}\right)^{-1}$, $I_0(x)$ is the modified Bessel function of the first kind and $E_0$ is the initial energy, i.e $P_0(E|E_0) = \delta(E-E_0)$. As expected, the equilibrium distribution   $P_\infty(E|E_0) = \beta\exp(-\beta E)$ does not depend on the initial conditions and is given by the Maxwell-Boltzmann distribution at temperature $\T{CM} = 1/(\kB \beta)$.
If the system is initially prepared at $t = 0$ in a steady state with energy distribution $P_0(E_0)$, the energy distribution after time $t$ is 

% =================================================
\begin{equation}
  P_E(E, t) = \int_0^\infty P_E(E|E_0, t)P_0(E_0)\d E_0.
\end{equation}
% =================================================

For an initial Maxwell-Boltzmann distribution, corresponding to a thermal equilibrium distribution at temperature $\T{init}$, the energy distribution at time $t$ is also a Maxwell-Boltzmann distribution

% =================================================
\begin{equation}
  P^\text{MB}_E(E, t) = \beta(t) e^{-\beta(t)E},
\end{equation}
% =================================================
with time dependent temperature 

% =================================================
\begin{equation}
  \T{CM}(t) = \T{\infty}+(\T{init}-\T{\infty})e^{-\g{CM} t}.
\end{equation}
% =================================================

Note that the initial temperature $\T{init}$ and final temperature $\T{\infty}$ can be controlled in the experiment by modulation of the trapping laser (feedback cooling), as discussed in section~\ref{sec:non-thermal} and demonstrated by Gieseler \emph{et al.} \cite{Gieseler2014}. Explicitly, a levitated nanoparticle can be cooled via feedback to a centre-of-mass temperature $\T{CM}$ far below the ambient temperature $\T{env}$. Once the feedback modulation is switched off, the particle will thermalize with the environment (in general via collisions with surrounding gas), at an average rate $\g{CM}$, which can be controlled by varying the gas pressure. 
The rate at which the particle relaxes to the new equilibrium state can also be accelerated by using time-dependent potentials \cite{Martinez2016}.
%Recent work has demonstrated the use of time-dependent potentials to accelerate the equilibration of a colloidal particle \cite{Martinez2016}.

\subsection{Fluctuation theorems}

As a system relaxes to a thermal equilibrium, the dynamics satisfy detailed balance with respect to the equilibrium distribution, and the time reversibility of the underlying dynamics implies that the transient fluctuation theorem

% =================================================
\begin{equation}
  \frac{P(-\dS)}{P(\dS)} = e^{-\dS},
\end{equation}
% =================================================

for the relative entropy change $\dS = \beta \Q+\Delta \entropy$ (or Kullback-Leibler divergence) holds. The quantity $\Delta \entropy = \entropy(t)-\entropy(0)$ is the difference in trajectory dependent entropy $\entropy(t) = -\ln P_0(u(t)) $ between the initial and the final states of the trajectory $u(t)$. The relative entropy change $\Delta \mathcal{S}$ is defined as the logarithmic ratio of the probability $P[u(t)]$
to observe a certain trajectory $u(t)$ and the probability $P[u^*(t)]$ of the time reversed trajectory $u^*(t)$, 

% =================================================
\begin{equation}
\Delta \mathcal{S} = \ln \frac{P[u(t)]}{P[u^*(t)]}.
\end{equation}
% =================================================

Here, $u(t)$ denotes an entire trajectory of length $t$ including position and momentum of the oscillator and
$u^*(t)$ denotes the trajectory that consists of the same states visited in reverse order with inverted momenta.
$\Q$ is the heat absorbed by the bath at reciprocal temperature $\beta$.
Because no work is done on the system, the heat $\Q$ exchanged along a trajectory equals the energy lost by the system, $\Q = -(E_t-E_0)$ where $E_0$ and $E_t$ are the energy at the beginning and at the end of the stochastic trajectory.
Note that the fluctuation theorem holds for any time $t$ at which $\dS $ is evaluated, and it is not required that the system has reached the equilibrium distribution at time $t$.

In general, the steady distribution $P_0(u(t))$ necessary to compute $\Delta\entropy$ is unknown. However, from the distribution derived for our model eqn.~\eqref{eq:Position_distribution_SS}, we find that for the relaxation from a non-equilibrium steady state generated by nonlinear feedback of strength $\fb$ and parametric modulation of strength $\emod$, the relative entropy change is given by

% =================================================
\begin{equation}
  \Delta \mathcal{S}=
 -\beta \frac{\emod\wo^2 \sin(2\phimod)}{2\Gamma\Omega}\left[E_t-E_0\right]
 -\beta \frac{\eta  \wo}{4m\Gamma \Omega^2}\left[E^2_t-E^2_0\right].
\end{equation}
% =================================================

Thus, our stochastic model allows us to express the relative entropy change during a relaxation trajectory in terms of the energy at the beginning and the end of that trajectory. This model was verified using a levitated nanoparticle by Gieseler \emph{et al.} \cite{Gieseler2014}, when starting from a variety of non-equilibrium steady states. 

%\todo{JG: move the following to intro?}
%This section illustrates the utility of the levitated nanoparticle system for studies of stochastic processes \cite{Ricci2017}, with the ability to control the potential and effective temperature through optical control, and the coupling to the bath through varying the pressure.

In addition to the fluctuation statistics during relaxation between steady states, one can also consider fluctuations during different protocols, e.g.  during a full thermodynamic cycle or while driving the particle with an external force $f(t)$ as was done by Hoang et al. \cite{Hoang:2018fn}, who verified a differential fluctuation theorem for the work $W = -\int_0^\tau   \dot{f}(t) q(t)   \d t$ 
% =================================================
\begin{equation}
  \frac{P\left(-W, u^*(t)\right)}{P\left(W, u(t)\right)} 
  = e^{-\beta(W-\Delta F)}.
\end{equation}
% =================================================

The differential fluctuation theorems can be integrated to yield a series of well known fluctuation theorems, such as the Jarzynski equality, the Crooks fluctuation theorem and the Hummer-Szabo relation. Thus, by verifying the underlying differential fluctuation theorem, the validity of the integral fluctuation theorem is implied. Importantly, the fluctuation theorems are valid for arbitrarily-far-from-equilibrium processes. Both detailed and integral fluctuation theorems allow the estimation of equilibrium free energy changes from nonequilibrium protocols and have found applications in determining the free energies of DNA molecules.
For a detailed review see Refs.~\cite{SeifertReview}.

\subsection{Heat Engines}
\label{sec:engines}
Technology is continuously miniaturizing, but as we pass below the micro-scale the challenge is not limited to the difficulty in constructing small devices. Once the work performed per duty cycle of an engine becomes comparable to the thermal energy of the piston, it is possible for the engine to run in reverse for short times, due to the fluctuating nature of energy transfer with the heat bath. This is exactly the scale at which biological systems operate, and a regime which levitated nano- and micro-particles have access to. 

To apply work to a trapped particle, one must either change (via a control parameter $\control(t)$) the trapping potential $U(q,\control)$, or apply an external force $f(q,\control)$, in which case the incremental work $\d\W$ reads:

% =================================================
\begin{equation}
\label{eqn:WI}
\d\W = (\partial U / \partial \control)\,\d\control + f\, \d q,
\end{equation}
% =================================================

with an associated heat increment:

% =================================================
\begin{equation}
\d\Q  = F\,\d q,
\end{equation}
% =================================================

where $F(q, \control) = -(\partial U)/(\partial q) +f$ is the total force acting on the particle, due to both the potential $U$ and the external force $f$. Importantly, the external force $f$ accounts for deterministic and stochastic contributions. Hence, along a trajectory $u(0) \to u(\tau)$:

% =================================================
\begin{eqnarray}
\label{eqn:totalWQ}
&\W(q(t)) = \int_0^\tau [ (\partial U / \partial \control) \dot{\control} + f\dot{q}]\,\d t \notag \\
&\Q(q(t)) = \int_0^\tau \dot{\Q}\,\d t = \int_0^\tau F \dot{q}\,\d t.
\end{eqnarray}
% =================================================

%%%%%%%%%%%%%%%%%%
\begin{figure}[b]
	\begin{center}
	{\includegraphics{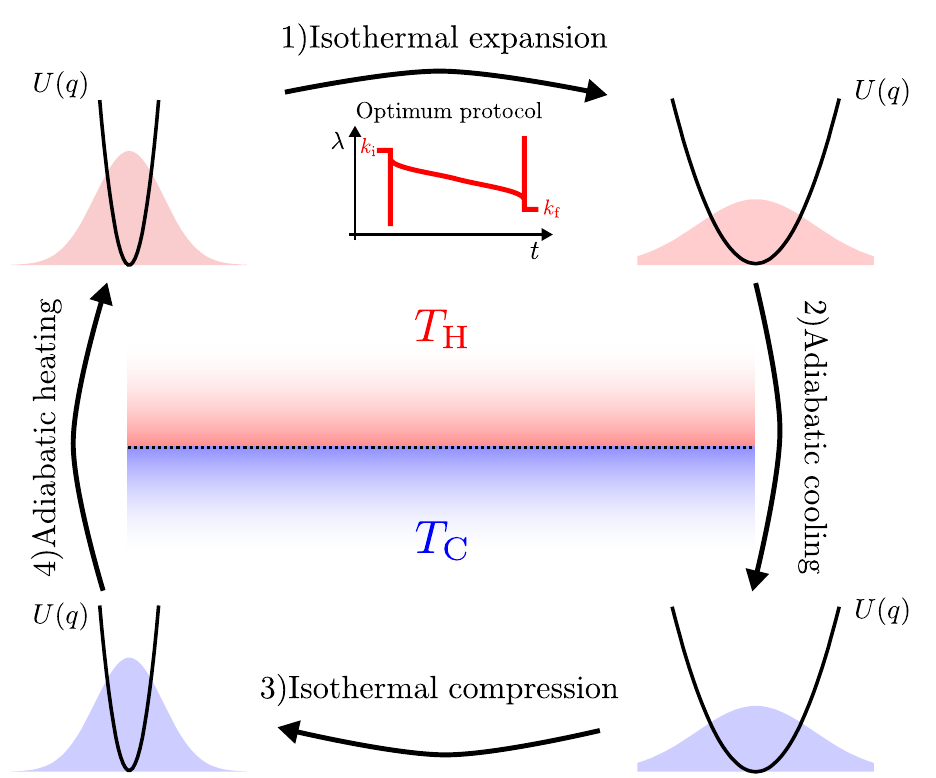}}
\caption{\label{fig:HeatEngine} \textbf{Stochastic heat engine.} This figure illustrates an engine cycle to realize a stochastic heat engine. A particle is confined by a potential $U(q,\control)$, and coupled to a hot/cold heat bath of temperature $\T{H/C}$. The shaded curves illustrate the position probability distribution of the particle. The cycle is explained in the text. The inset below step 1) illustrates an example optimum protocol for realising the compression step for an underdamped heat engine. The trap is expanded by lowering the spring constant from a value $\ks{i}$ to $\ks{f}$, with the most efficient protocol involving sharp parameter variations (adapted from \cite{Dechant2014}).}
\end{center}
\end{figure}
%%%%%%%%%%%%%%%%%%%

Next, we will explicitly apply this to the case of a heat engine. We discuss heat engines since they are an extremely useful machine. An engine, or motor, converts one type of energy into mechanical work, and a heat engine specifically converts heat flow between two reservoirs into mechanical work, particularly useful since heat is often generated as a waste product. Schmiedl \& Seifert gave the first full description of a colloidal stochastic heat engine \cite{Schmiedl2008}, which was realized by Blickle \& Bechinger a few years later \cite{Blickle2011}. The engine operates under the following cyclic process (as illustrated in fig.~\ref{fig:HeatEngine}): 1) an isothermal transition at temperature $\T{H}$ with a time dependent variation of $U(q,t)$ to extract work $\W<0$; 2) an instantaneous reduction in temperature $\T{H}\to \T{C}$, where no heat is exchanged with the bath (adiabatic); 3) an isothermal transition at $\T{C}$ with a time dependent variation of $U(q,t)$ and $\W>0$; 4) an instantaneous increase in temperature $\T{C} \to \T{H}$. 

For a harmonically confined particle $U(q,t) = \ks{}(t)q(t)^2/2$, where $\ks{}$ is the trap stiffness, our control parameter $\control(t) \equiv \ks{}(t)$. Other choices of $\control$ could include a movement of the trap centre. Reducing $\ks{}$ corresponds to an expansion $\langle W\rangle <0$, as the confinement is weakened. Following references \cite{Schmiedl2008, Dechant2017}, it is convenient to analyse this scenario by considering the equations of motion for the variance $\var{q}(t) \equiv \langle q^2(t) \rangle$, with the equation of motion:

% =================================================
\begin{equation}
\label{eqn:sigmax}
\dvar{q} = -m\mu\ddvar{q} -2\mu\ks{}(t)\var{q} +2m\mu \var{v}, 
\end{equation}
% =================================================

where $\mu = 1/(m\g{CM})$, $\var{v}(t) \equiv \langle \dot{q}^2(t) \rangle$, and also noting for the harmonic oscillator that the frequency $\w(t)= \sqrt{\ks{}(t)/m}$. This can be simplified in the overdamped regime $\g{CM} >> \w$, first by removing the inertial term $\propto \ddvar{q}$, and secondly by assuming that the state is always thermal $\var{v} = \kB \T{CM}/m$ \cite{Dechant2017}. This yields the overdamped equation of motion,

% =================================================
\begin{equation}
\label{eqn:sigmax_od}
\dvar{q} = -2\mu \ks{}(t)\var{q} + 2\mu \kB{} \T{CM}. 
\end{equation}
% =================================================

 Using eqn.~\eqref{eqn:totalWQ} we find for the total work $\W$ along an isothermal trajectory at $\T{CM}$ from time $t_{\rm i} \to t_{\rm f}$,

% =================================================
\begin{equation}
\W(\ks{}(t)) = \int_{t_{\rm i}}^{t_{\rm f}} \dot{\ks{}}\frac{\var{q}}{2}\,\d t, 
\end{equation}
% =================================================

where it is evident that the work done on the particle depends on the rate at which the potential is changed. For an instantaneous change in spring constant, where the position distribution of the particle does not have time to change, the work done is simply $\Delta \W = \frac{1}{2}[\ks{}(t_{\rm f}) - \ks{}(t_{\rm i})]q(t_{\rm i})^2$. More generally, through solving eqn.\ \eqref{eqn:sigmax_od} for $\ks{}(t)$ one finds the full expression for the work along the trajectory:

% =================================================
\begin{equation}
\W(\ks{}(t)) = \frac{1}{4\mu}\int_{t_{\rm i}}^{t_{\rm f}} \frac{(\dvar{q})^2}{\var{q}} - \frac{1}{2}\T{CM}[\ln\var{x}]_{t_{\rm i}}^{t_{\rm f}} + \frac{1}{2}[\ks{}(t)\var{q}]_{t_{\rm i}}^{t_{\rm f}}. 
\end{equation}
% =================================================

Hence, using this expression, by monitoring the motion of a colloidal particle as it undergoes the cyclic heat engine, one can extract the work statistics. We leave a full discussion of the heat and entropy statistics to other sources, for example Spinney \& Ford \cite{FordBook}. 

% The first realization of the colloidal heat engine was by Blickle \& Bechinger \cite{Blickle2011}, who locally heated the liquid (water) surrounding their trapped particle through laser absorption, realising temperature changes of $70^\circ$C in 10\,ms, and their data agreed well with theoretical predictions. If we consider the simplified case of an instantaneous change in $\ks{}$, such that $\Delta \ks{} \equiv \ks{f} - \ks{i}$, and follow the cycle in Fig.~\ref{fig:HeatEngine}, then the mean work done in a cycle is $\langle \W \rangle = \frac{1}{2} \Delta \ks{} [\langle q^2 \rangle_{\rm C} - \langle q^2 \rangle_{\rm H}]$, where the subscripts C, H refer to the values in the cold and hot baths respectively. Hence, since for a harmonic oscillator $\langle q^2 \rangle \propto \T{CM}$, in this cycle work is extracted from the particle $\langle W \rangle < 0$. However, Blickle \& Bechinger observed that for \emph{individual} cycles sometimes $\Delta \W > 0$, due to the fluctuating position statistics. 

How does this discussion of heat engines change in the underdamped regime? In eqn.~\eqref{eqn:sigmax_od} we simplified the equation of motion in the overdamped regime, such that the position was independent of the velocity. This simplification allows one to analytically construct protocols (the way in which $\control$ changes over time) that maximize the efficiency of a stochastic heat engine \cite{Schmiedl2008}. The overdamped efficiency of a microscopic heat engine can even exceed the Curzon-Ahlborn efficiency-at-maximum-power limit $\eta^* = 1-\sqrt{\T{C}/\T{H}}$ for macroscopic engines \cite{Schmiedl2008}. 

An analytic solution is not known in the underdamped case, where the position and velocity variables cannot be separated, and numerical methods must be used, which find that the efficiency of the underdamped stochastic heat engine is bounded by $\eta^*$  \cite{GomezMarin2008}. In both regimes, the optimum protocols call for instantaneous jumps in the control parameter\footnote{Instantaneous changes minimize work dissipation, since they minimize the time over which a particle is accelerating \cite{GomezMarin2008}.} $\control$, as illustrated in the inset to fig.~\ref{fig:HeatEngine}. In the overdamped regime, a particle reacts slowly to changes in $\control$, whereas in the underdamped regime it reacts rapidly. Hence, although in theory the overdamped efficiency may be higher, practically it may be easier to realize optimum work extraction cycles with an underdamped engine.

To realize an underdamped heat engine, one has to engineer a coupling to an effective heat bath (since by definition an underdamped system is weakly coupled to the environment). Such a coupling is described in detail in section~\ref{sec:effective_potential}. Dechant \emph{et al.} \cite{Dechant2014} propose to realize an underdamped heat engine through a combination of optical cavity cooling and thermalization with residual gas. Another option would be to levitate a charged particle in a Paul trap and provide the heat bath via noise applied to nearby electrodes \cite{Martinez2015, Goldwater2018}, which may be more suitable for operation in the quantum regime.

\section{Acknowledgements}

(JM): This project is supported by the European Research Council (ERC) under the European Union’s Horizon 2020 research and innovation programme (Grant agreement No. 803277), and by EPSRC New Investigator Award EP/S004777/1. \\
(JG): This project has received funding from the European Union’s Horizon 2020 research and innovation programme under the Marie Skłodowska-Curie grant agreement No. 655369.

\bigskip

\bibliographystyle{unsrt}
\bibliography{Thermo}

%\bibliographystyle{plain}
%\bibliography{Bibliography}
%\nobibliography{}

%%%
%\begin{thebibliography}{99}
%
%\bibitem{Mukamel03}
%S. Mukamel, {\it Phys. Rev. Lett. } {\bf 90}, 170604 (2003). 
%
%\end{thebibliography}
%%%

\end{document}

%% file: commands.tex
\newcommand{\mum}{\mu \text{m}} % micrometers
\newcommand{\mbar}{\text{mbar}} % milli Bars
\newcommand{\Hz}{\text{Hz}} % Hertz
\newcommand{\pN}{\text{pN}} % PicoNewtowns

\renewcommand{\d}{{\rm d}}% differential d
\newcommand{\kB}{{\rm k_B}}% Boltzmann's constant
\newcommand{\waist}{\text{w}}% beam waist of Gaussian beam
\newcommand{\w}{\Omega}% mechanical frequency
\newcommand{\wo}{\Omega_0}% mechanical resonance frequency
\newcommand{\ws}{\Omega_s}% mechanical resonance during squeezing pulse

\newcommand{\Kn}{{\rm Kn}}% Knudsen number
\newcommand{\Pg}{P_{\rm gas}}% gas pressure
\newcommand{\dm}{{\rm d_m}}% gas molecule diameter
\newcommand{\cg}{{\rm c_{\rm P}}}% coefficient for gas pressure in linear regime
\newcommand{\crosssectiongas}{\sigma_\text{gas}}% cross section of gas
\newcommand{\viscosity}{\mu_v}
\newcommand{\mgas}{m_\text{gas}}% mass of gas molecules
\newcommand{\eo}{\epsilon_0}% vacuum permittivity
\newcommand{\er}{\epsilon_{\rm r}}% relative permittivity
\newcommand{\eom}[1]{{\epsilon(\omega_{\rm #1})}} % Frequency dependent permittivity
\newcommand{\acccoeff}{c_{\rm acc}}% accommodation coefficient 
\newcommand{\hcap}{c_{\rm HC}} %heat capacity
\newcommand{\vth}{v_{\rm th}} % gas thermal velocity
\newcommand{\whn}{\Xi} % Random noise

\newcommand{\rodd}{d}
\newcommand{\rodl}{l}

\newcommand{\geff}{\tilde{\Gamma}}% damping coefficient of effective Brownian particle (driven potentials)
\newcommand{\Ueff}{U_\text{eff}}% effective potential of driven particle

\newcommand{\g}[1]{\Gamma_{\rm #1}}% damping coefficient
\newcommand{\T}[1]{T_{{\rm #1}}}% Temperature
\newcommand{\dpcoeff}{c_{\rm dp}}% dipole coefficient
\newcommand{\Pscat}{P_{\rm scat}}% scattered power
\newcommand{\Popt}{P_{\rm opt}}% optical  power
\newcommand{\Poptm}{\bar{P}_{\rm opt}}% mean optical  power
\newcommand{\Iopt}{I_{\rm opt}}% optical intensity
\newcommand{\Uopt}{U_{\rm opt}}% optical intensity

\newcommand{\wopt}{\omega_{\rm L}}% optical frequency
\newcommand{\kopt}{k_{\rm L}}% optical wavevector
\newcommand{\lambdaopt}{\lambda_{\rm L}}% optical wavelength
\newcommand{\polar}[1]{\alpha_{\rm #1}}% polarisability
\newcommand{\scatcross}{\sigma_{\rm scat}}% scattering cross section
\newcommand{\totcross}{\sigma_{\rm tot}}% total cross section
\newcommand{\abscross}{\sigma_{\rm abs}}% absorption cross section
\newcommand{\F}[1]{\mathbf{F}_{\rm #1}}% force
\newcommand{\Ffluct}{\mathcal{\boldsymbol{\mathcal{F}}}_{\rm fluct}}% stochastic force
\newcommand{\omL}{\omega_{\rm L}} % Laser frequency
\newcommand{\ombb}{\omega_{\rm bb}} % frequency of blackbody radiation
\newcommand{\ks}[1]{{\rm k_{\rm #1}}} % Spring constant

\newcommand{\Qf}{{\rm Q}} % Quality factor

\newcommand{\Sff}[1]{S_{\rm ff}^{\rm #1}}% force spectral density
\newcommand{\noise}{n}% index for noise contributions

\newcommand{\chitensor}{\boldsymbol{\chi}}

\newcommand{\depol}{\boldsymbol{N}}

\newcommand{\q}{q}% coupling coefficient to external field

\newcommand{\var}[1]{\sigma_{#1}^2}% variance
\newcommand{\dvar}[1]{\dot{\sigma}_{#1}^2}% dot variance
\newcommand{\ddvar}[1]{\ddot{\sigma}_{#1}^2}% double dot variance
\newcommand{\corr}[2]{\langle #1(t)#2(0) \rangle}% correlation function

\renewcommand{\a}{A} % oscillation amplitude
\newcommand{\sqrtE}{\varepsilon} % square root of energy

\newcommand{\R}[1]{{\rm R}_{#1}} % rates in Kramers theory 

\newcommand{\emod}{\zeta} % modulation depth
\newcommand{\phimod}{\phi_{\rm mod}} % modulation phase
\newcommand{\omegamod}{\Omega_{\rm mod}} % modulation frequency
\newcommand{\fb}{\eta} % feedback coefficient

\newcommand{\dS}{\Delta\mathcal{S}} % relative entropy change
\newcommand{\Q}{\mathcal{Q}} % heat 
\newcommand{\W}{\mathcal{W}} % work
\newcommand{\entropy}{\Phi} % entropy

\newcommand{\diag}{\text{diag}}
\newcommand{\polarangle}{\phi}
\newcommand{\azimuthalangle}{\theta}

\newcommand{\1}{\mathbb{I}}

\newcommand{\control}{\lambda_c}